\documentclass[usenatbib]{mn2e}
\usepackage{natbib}
\usepackage{amsmath,amssymb}
\usepackage{epsfig}
\usepackage{color}

\input{macros.tex}

\title[Constraining the Galaxy's dark halo with RAVE stars]
{Constraining the Galaxy's dark halo with RAVE stars}

\author[T. Piffl et al.]{T.~Piffl$^1$\thanks{E-mail: tilmann.piffl@physics.ox.ac.uk},
J.~Binney$^1$,
P.~J.~McMillan$^1$,
M.~Steinmetz$^{2}$,
A.~Helmi$^3$,
R.~F.~G.~Wyse$^4$,
\newauthor
O.~Bienaym\'{e}$^5$,
J.~{Bland-Hawthorn}$^6$,
K.~Freeman$^7$,
B.~Gibson$^{8,9}$,
G.~Gilmore$^{10}$,
\newauthor
E.~K.~Grebel$^{11}$,
G.~Kordopatis$^{10}$,
J.~F.~Navarro$^{12}$,
Q.~Parker$^{13,14,15}$,
\newauthor
W.~A.~Reid$^{13,14}$,
G.~Seabroke$^{16}$,
A.~Siebert$^5$,
\newauthor
F.~Watson$^{15}$,
T.~Zwitter$^{17,18}$
\vspace*{3ex}
\\
$^1$Rudolf Peierls Centre for Theoretical Physics, Keble Road, Oxford OX1 3NP, UK\\
$^2$Leibniz-Institut f\"ur Astrophysik Potsdam (AIP), An der Sternwarte 16, 14482 Potsdam, Germany\\
$^3$Kapteyn Astronomical Institute, University of Groningen, P.O. Box 800, 9700 AV Groningen, The Netherlands\\
$^4$Department of Physics \& Astronomy, Johns Hopkins University, Baltimore, MD 21218, USA\\
$^5$Observatoire astronomique de Strasbourg, UMR 7550 CNRS/Universit\'{e} de Strasbourg, Strasbourg, France\\
$^6$Sydney Institute for Astronomy, University of Sydney, NSW 2006, Australia\\
$^7$Research School of Astronomy and Astrophysics, Australian National University, Cotter Rd., Weston, ACT 2611, Australia\\
$^8$Institute for Computational Astrophysics, Dept of Astronomy \& Physics, Saint Mary's University, Halifax, NS, BH3 3C3, Canada\\
$^9$Jeremiah Horrocks Institute, University of Central Lancashire, Preston, PR1 2HE, UK\\
$^{10}$Institute for Astronomy, University of Cambridge, Madingley Road, Cambridge CB3 0HA, UK\\
$^{11}$Astronomisches Rechen-Institut, Zentrum f\"ur Astronomie der Universit\"at Heidelberg, M\"onchhofstr.\ 12--14, 69120 Heidelberg,
Germany\\
$^{12}$Senior CIfAR Fellow, University of Victoria, Victoria BC, Canada V8P 5C2\\
$^{13}$Department of Physics, Macquarie University, Sydney, NSW 2109, Australia\\
$^{14}$Research Centre for Astronomy, Astrophysics and Astrophotonics, Macquarie University, Sydney, NSW 2109, Australia\\
$^{15}$Australian Astronomical Observatory, PO Box 915, North Ryde, NSW 1670, Australia\\
$^{16}$Mullard Space Science Laboratory, University College London, Holmbury St Mary, Dorking, RH5 6NT, UK\\
$^{17}$University of Ljubljana, Faculty of Mathematics and Physics, Jadranska 19, Ljubljana, Slovenia\\
$^{18}$Center of excellence space-si, Askerceva 12, Ljubljana, Slovenia
}

\begin{document}

\date{Accepted 2014 September 16.  Received 2014 September 11; in original form 2014 June 14}

\pagerange{\pageref{firstpage}--\pageref{lastpage}} \pubyear{2014}

\maketitle

\label{firstpage}

\begin{abstract}
We use the kinematics of $\sim200\,000$ giant stars that lie within
$\sim1.5\kpc$ of the plane to measure the vertical profile of mass density
near the Sun.  We find that the dark mass contained within the isodensity surface
of the dark halo that passes through the Sun ($(6\pm0.9)\times10^{10}\msun$),
and the surface density within $0.9\kpc$ of the plane
($(69\pm10)\msun\pc^{-2}$) are almost independent
of the (oblate) halo's axis ratio $q$. If the halo is spherical, 46 per cent of the
radial force on the Sun is provided by baryons, and only 4.3 per cent of the
Galaxy's mass is baryonic.  If the halo is flattened, the baryons contribute
even less strongly to the local radial force and to the Galaxy's mass.
The dark-matter density at the location of the Sun is
$0.0126\,q^{-0.89}\msun\pc^{-3}=0.48\,q^{-0.89}\hbox{\,GeV}\cm^{-3}$. When combined
with other literature results we find hints for a mildly oblate dark halo
with $q\simeq0.8$.
Our value for the dark mass
within the solar radius is larger than that predicted by cosmological
dark-matter-only simulations but in good agreement with simulations once
the effects of baryonic infall are taken into account.
Our mass models consist of three double-exponential discs, an oblate bulge and a
Navarro-Frenk-White dark-matter halo, and we model the dynamics of the RAVE
stars in the corresponding gravitational fields by finding distribution
functions $f(\vJ)$ that depend on three action integrals.  Statistical errors
are completely swamped by systematic uncertainties, the most important of
which are the distance to the stars in the photometric and spectroscopic
samples and the solar distance to the Galactic centre.  Systematics other than the flattening of the dark halo yield
overall uncertainties $\sim15$ per cent.
\end{abstract}

\section{Introduction}

There is abundant evidence that halos of still undetected dark matter
dominate the mass budgets of galaxies. Galaxies with luminosities similar to
that of the Milky Way have the largest baryon fractions, yet even in these
systems we think there is at least an order of magnitude more dark matter
than stars and gas. This belief is founded on numerous lines of evidence, but
two important and essentially independent ones are the dynamics of the Local
Group \citep{Kahn1959, Li2008} and fits of $\Lambda$CDM cosmologies to the
cosmic background radiation combined with numerical simulations of galaxy
formation \citep[e.g.][]{Bower2010}.

Since attempts to detect dark matter through its weak interactions in
underground experiments have yet to bear fruit \citep{Mirabolfathi}, dark matter is so far only
detectable through its gravitational field. Studies of weak lensing
\citep{Velander2014} and thermal X-ray emission \citep{Das2011} have been used to map dark
matter on large scales, while the kinematics of stars and gas provide the
most powerful probe on small scales. For intermediate regions strong lensing can provide some insights \citep[e.g][]{Kneib2011}..

Studies of star and gas kinematics in our own Milky Way have to date
constrained the dark-matter distribution less strongly than have similar
studies in external galaxies, such as NGC\,3198 \citep{vanAlbada1985}. The
cause of this surprising fact is essentially that dark matter is most
important at large radii, where it becomes difficult to determine distances
to objects with measured velocities. Moreover, the Sun's velocity and location with
respect to the Galactic centre has an uncertainty of several percent, and
this uncertainty propagates to the Galactocentric velocities of all tracers,
which are inevitably measured in the heliocentric frame.

It is thought that of order half the inward force on the Sun derives from
dark matter and half from stars and gas, predominantly in the disc \citep{Sackett1997}. A
significant step towards testing this conviction would be to show that the
structure of the Galaxy's gravitational field perpendicular to the plane is
consistent with roughly half the mass interior to the Sun being contained in
the disc and the other half contained in a roughly spherical halo \citep[e.g.][]{Sackett1997}. One of
the original goals of the RAdial Velocity Experiment \citep[RAVE]{RAVE_DR1}
was to test this hypothesis by gathering spectra of large numbers of stars
within the ``extended solar neighbourhood'', the region within $\sim2\kpc$ of
the Sun.
In this article we describe our attempt in this direction. The independent parallel work by \citet{Bienayme2014} has similar goals, but a very different methodology. Recently, also the SEGUE survey \citep{SEGUEpaper} was used by \citet{Zhang2013} to estimate the local dark matter density.

The fourth data release from RAVE gives line-of-sight velocities and stellar
parameters for $\sim400\,000$ stars in the extended solar neighbourhood
\citep{RAVE_DR4}. Using these stellar parameters, \citet{Binney2014a}
estimated distances to the stars, which are roughly half giants and half
dwarfs. Combining the RAVE data with proper motions from the UCAC4
\citep{UCAC4_paper}, \citet{Binney2014b} derived the velocity  distributions
of stars in eight spatial bins, four inside the solar radius $R_0$
and four outside it, and at various distances from the plane. They showed
that these velocity distributions were in remarkably good agreement with
those predicted by a dynamical model of the galaxy that \citet[][hereafter
B12b]{Binney2012b} had fitted to the velocity distribution at the Sun
determined by the Geneva--Copenhagen survey \citep[hereafter
GCS]{GCS2004,Holmberg2007} and the estimate of density versus distance from
the plane derived by \citet{Gilmore1983}. 

Although the agreement between data and model was near perfect near the
plane, at distances $|z|\gta0.5\kpc$ the model failed to reproduce the data
in two respects: (i) the distribution of the components $V_1$ of velocity
parallel to the longest principal axis of the velocity ellipsoid (which
points near to the Galactic Centre) was predicted to be too narrow; (ii) the
predicted distribution of azimuthal components $V_\phi$ tended to be slightly
displaced to small $V_\phi$ with respect to the observed distribution.
Neither defect is an inherent feature of an axisymmetric equilibrium model.

In this paper we have two goals: (i) to seek a modified form of the
distribution function (\df) of the B12b model that is consistent with the
RAVE data, and (ii) to use models obtained in this way to constrain the local
gravitational field and thus the distribution of dark matter. 
\section{Data} \label{sec:Data}
In this section we introduce the data we have used and explain how we
exploited them to constrain our Galaxy model. In order to make model
predictions for these measurements we assume a distance of the Sun to the
Galactic centre (GC), $R_0$, to be 8.3 kpc \citep[e.g.][]{Gillessen2009b,
McMillan2011, Schoenrich2012}, the position of the Sun above the Galactic
plane, $z_0$, to be 14 pc \citep{BinneyGS1997} and the solar motion with
respect to the local standard-of-rest (LSR), $\vv_\odot$, to be
$(11.1,12.24,7.25)\kms$ \citep{Schoenrich2010}.
\subsection{Gas terminal velocities}
The distribution of \HI\ and CO emission in the longitude-velocity plane yield
a characteristic maximum (``terminal'') velocity for each line of sight
\citep[e.g.][\S9.1.1]{GalacticAstronomy}. The terminal velocities are related
to the circular speed $v_{\rm c}(R)$ by
\begin{equation}
 \begin{split}
  v_\mathrm{term}(l) &= {\rm sign}(\sin l) v_\mathrm{c}(R) - v_\mathrm{c}(R_0)\sin l \\
                     &= {\rm sign}(\sin l) v_\mathrm{c}(R_0|\sin l|) - v_\mathrm{c}(R_0)\sin l.
 \end{split}
\end{equation}
We use the terminal velocities $v_{\rm term}(l)$ from \citet{Malhotra1995}.
Following \citet{Dehnen1998} and \citet{McMillan2011} we neglect data at
$\sin l < 0.5$ in order not to be influenced by the Galactic bar, and we
assume that the ISM has a Gaussian velocity distribution of dispersion
7~km\,s$^{-1}$.
\subsection{Maser observations}
\citet{Reid2014} presented a compilation of 103 maser observations that
provide precise 6D phase space information. Since masers are associated with
young stars their motions are very close to circular around the GC. We again
assume an intrinsic velocity dispersion of 7~km\,s$^{-1}$ and no lag against
the circular speed \citep{McMillan2010}. For the likelihood computation we
neglected 15 sources that were flagged as outliers by \citet{Reid2014} and
also all sources at $R < 4\kpc$. The latter is again to prevent a bias by the
Galactic bar. To assess the likelihood of a maser observation, we predict the
observed velocities (line-of-sight velocity, proper motions) as functions of
heliocentric distance and then integrate the resulting probability density
along the line-of-sight.
\subsection{Proper motion of SgrA*}
\citet{Reid2004} measured the proper motion of the radio source SgrA* in the
GC to be
$$
 \mu_\mathrm{SgrA^\star} = -6.379 \pm 0.024~\mathrm{mas\,yr}^{-1}.
$$
This source is thought to be associated with the super-massive black hole that
sits in the gravitational centre of the Milky Way with a velocity below
$1\kms$. Hence this measurement reflects the solar motion with respect to
the GC.
\subsection{Mass within 50 kpc}
By modelling the kinematics of the Galaxy's satellites, \citet{Wilkinson1999}
estimated the mass of the Galaxy within $50\kpc$ to be
$$
 M(r<50\kpc) = 5.4^{+0.2}_{-3.2} \times 10^{11}\msun
$$
Following \citet{McMillan2011}, we use this measurement as an upper limit on
the mass of our models. This limit effectively sets the scale radius
$r_{\rm 0,dm}$ of the halo because all plausible models have $M(r<50\kpc)$
very close to the Wilkinson \& Evans' upper limit.
\subsection{Concentration of the dark halo} \label{sec:concentration_prior}
The dark halo has two free parameters: the normalising density
$\rho_\mathrm{0,dm}$ and the scale radius $r_\mathrm{0,dm}$. We constrain
the scale radius using the most likely value for the concentration parameter
$c$ as found in cosmological dark-matter-only simulations. The normalising
density of the dark halo is
\begin{equation}
 \rho_\mathrm{0,dm} = \frac{3H^2}{8\pi G}\delta_c,
\end{equation}
 where $\delta_c$ is related to the halo's concentration parameter $c$ by
\begin{equation}
  \delta_c = \frac{\Delta_\mathrm{vir}}{3}\frac{c^3}{\ln(1+c) - c/(1+c)} ,
\end{equation}
with the parameter $\Delta_\mathrm{vir}$ being the virial over-density.
We take the Hubble constant to be $H=73\kms$\,s$^{-1}$\,Mpc$^{-1}$, and from
\citet{BoylanKolchin2010} adopt
\begin{equation}
\begin{split}
\Delta_\mathrm{vir}&=94\\
 \ln c&= 2.256 \pm 0.272.
\end{split}
\end{equation}
With these choices, the local dark-matter density $\rho_\mathrm{dm}(R_0,z_0)$
becomes a function of only the dark-halo's scale radius $r_{\rm 0,dm}$, so
the latter can be  determined once we have chosen $\rho_\mathrm{dm}(R_0,z_0)$.
\subsubsection{Baryon fraction}
By matching the number density of SDSS galaxies with different stellar masses
to that of dark-matter halos in the Millennium simulations, \citet{Guo2010}
obtained a relation between stellar mass $M_\star$ and $M_{200}$, the mass
interior to a Galactocentric sphere within which the mean density 200 times
the critical density for closure of the Universe. Their relation is
\begin{equation}
 M_\star = M_{200} \times A\left[\left(\frac{M_{200}}{M_0}\right)^{-\alpha}
             + \left(\frac{M_{200}}{M_0}\right)^{\beta}\right]^{-\gamma},
\end{equation}
where $M_0 = 10^{11.4}\msun$, $A=0.129$, $\alpha = 0.926$, $\beta = 0.261$
and $\gamma = 2.440$. We have imposed this constraint with uncertainty 0.2
in $\log_{10}M_\star$.
\subsection{The RAVE survey}
The RAVE survey has taken spectra at resolution $R\simeq7500$ of
$\sim500\,000$ stars that have 2MASS photometry. Stars were selected for
observation based on their $I$ band apparent magnitudes; $I\approx9-13$.
Stellar parameters were extracted from the spectra by a pipeline described
in \citet{RAVE_DR4}, and from those parameters and the 2MASS photometry
\citet{Binney2014a} determined probability density functions (pdfs) for the
distances of most stars. Roughly half the stars are giants, and these stars
form our main sample. Within the survey's observing cone, the giants sample
densely the region within $\sim2\kpc$ of the Sun, so they are ideally suited
for determination of the vertical structure of the Galaxy's disc.
\citet{Binney2014b} examine the kinematics of these stars in some detail.

We clean the data using the following criteria:
\begin{itemize}
 \item[--] Signal-to-Noise ratio (S/N) $>10$.
 \item[--] For stars with multiple observations we chose the observation with the
 highest S/N.
 \item[--] The stars have a parallax estimate in \citet{Binney2014a}.
 \item[--] Proper motion uncertainty $<8\mas\yr^{-1}$.
 \item[--] $|U|$ and $|W|$ velocity components $< 350\kms$.
 \item[--] The stars lie in the cylindrical shell around the Galactic centre
 $|R-R_0|<1\kpc$.
\end{itemize}
The restrictions in the $U$ and $W$ velocities were introduced to weed out
unreliable proper motion and distance estimates. The last criterion selects for
relatively nearby stars in RAVE for which we have more trust in the distance
estimations. Note, that throughout this study we do not look at the azimuthal
positions of the stars w.r.t.\ the Galactic centre and hence implicitly assume
axis-symmetry (as we will do in our Galaxy model). This is, of course, an
approximation, as there is known sub-structure in the local stellar population
\citep{Dehnen1998_UVplane, Siebert2012, Antoja2012, Williams2013}.

We then define a sample of giant stars through
$$
 \logg < 3.5\,\dex\,.
$$
This leaves us with a sample of 181\,621 stars. For later use we also define a
sample of hot dwarf stars using the following cuts:
$$
\begin{array}{rcl}
 \logg &>& 3.5\,\dex, \\
 \Teff &>& 6000\K.
\end{array}
$$
Here we obtain a sample of 55\,398 stars.
\subsection{The vertical stellar density profile}
Any dynamical mass measurement requires knowledge of the spatial distribution
of a tracer population in addition to knowledge of that population's
kinematics. Hence we need an estimate of the spatial distribution of the
population from which RAVE stars are drawn. The seminal study of
\citet{Gilmore1983} provided an estimate of the vertical density profile of
dwarf stars that remains valuable even though it is based on observations of
only $\sim2700$ objects\footnote{Their estimate for stars with $M_V \in [3,4]$.}.
More recent data sources include the 2MASS catalogue and the Sloan Digital
Sky Survey \citep[SDSS,][]{York2000}. In a subsequent paper we will constrain
the Galaxy's mass distribution by combining RAVE data with star counts from
both 2MASS and SDSS: use of both surveys is desirable because the magnitude
limit of 2MASS is too bright to give sensitivity to the thick disc, and the
magnitude limit of SDSS is too faint to give sensitivity to the thin disc.

However, here we adopt a less sophisticated approach based on the vertical
density profile that \citet[][hereafter \citetJuric]{Juric2008} deduced from
the SDSS survey through a main-sequence colour-magnitude relation. We use the
data points shown in the middle panel of their Figure~15, which shows results
from M dwarf stars in the colour range $0.70 < r-i < 0.80$. Similar to RAVE,
this sample should carry only a weak metallicity bias and none in age.
\citetJuric\ fitted their data with the density profile
\[
\nu(R=R_0,z)\propto\exp(-|z|/\zthin)+\fthick\exp(-|z|/\zthick).
\]
 and used artificial star tests to estimate the impact of Malmquist
bias and stellar multiplicity on the fitted density profile. They concluded
that the fitted scale heights of both the thin and the thick disc have to be
increased 5 + 15 = 20 per cent to compensate for the two effects, while the
parameter $\fthick$ that sets the relative importance of the discs,
should be decreased by 10 per cent. This results in $\zthin = 0.3\kpc$,
$\zthick = 0.9\kpc$ and $\fthick = 0.12$.

Here Malmquist bias refers to the fact that in the presence of random errors
in  the apparent and/or the absolute magnitudes of stars
-- i.e.\ effectively in their distance moduli $\mu$ -- their distance
estimates will be on average too small. The stars in \citetJuric\ have
magnitude-dependent photometric uncertainties of 0.02--0.12~dex and there is
an uncertainty in the estimates of the absolute magnitude coming from the
finite width of the main sequence that is not covered by J08's
colour-magnitude relation.

\begin{figure}
 \centering
 \includegraphics[width=0.47\textwidth]{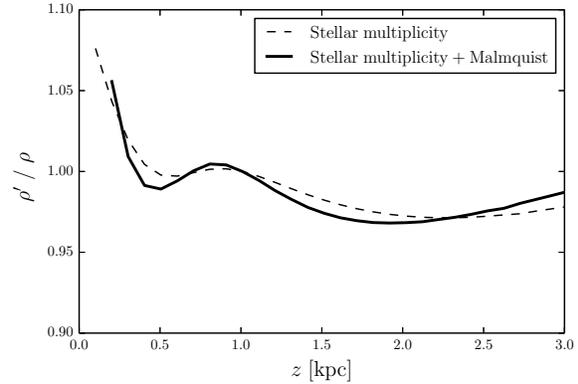}
 \caption{Ratio of the artificially biased vertical mass density profile
 $\rho'$ to the original profile $\rho$ as a function of height above the
 Galactic plane $z$.}
 \label{fig:bias-induction}
\end{figure}

Before we compare the data to our predictions, we have
to introduce these biases into the model. Malmquist bias is modelled by
folding the vertical stellar number count profile $N(z) \propto 2\pi\,z^2
\nu(R_0,z)$ with a Gaussian of width 0.32~dex in distance modulus
$\mu$. This width is a combination (in quadrature) of the maximal measurement
uncertainty of 0.12~dex and a value 0.3~dex for the finite width of the main
sequence. For stellar multiplicity the modelling is more uncertain. We use
the simplifying assumptions that (1) we have only binaries and (2) that the
companions have about the same brightness as the primaries as such systems
occur preferentially \citep{Delfosse2004} and also produce most of the
effect. This results in an underestimate of their true distance by a factor
of $1/\sqrt{2}$. The resulting new profile $N'(z)$ is then computed from the
original profile $N(z)$ as follows
\begin{equation}
 N'(z) = (1-f_{\rm binary}) N(z) + f_{\rm binary} N(\sqrt{2}z),
\end{equation}
where $f_{\rm binary}$ is the binary fraction. The biased number count
profile is then converted back into a density profile
$\propto N'(z)/(2\pi\,z^2)$. \citet{Dieterich2012} report a fraction of about
10 per cent of M dwarfs having M dwarfs companions and we adopt this number.
\figref{fig:bias-induction} illustrates the effect of the two biases. Note,
that \citetJuric\ assumed higher binary fraction of 35 per cent to correct their fit
results. This led to more significant corrections than those applied here.
The differences are well within the range of systematic uncertainties reported below.
%
%%%%%%%%%%%%%%%%%%%%%%%%%%%%%%%%%%%%%%%%%%%%%%%%%%%%%%%%%%%%%%%%%%%%%%%%%%%%%%%
%
%
\section{Methodology} \label{sec:methodology}
For a number of candidate mass distributions and associated gravitational
potentials we choose the parameters of a \df\ for the Galaxy's stars such that
the \df\ correctly predicts the distributions of the three principal velocity
components in each of eight spatial bins around the Sun. The selected \df\
then predicts the vertical density profile of stars in the solar cylinder.
If the mass model assigns most mass to the disc, the predicted stellar
profile decreases very steeply with distance from the plane because the
gravitational potential has a deep minimum at $z=0$. Conversely, if the mass
model assigns little mass to the disc, the potential's minimum at $z=0$ is
shallow and the predicted stellar density declines slowly with distance from
the plane. We identify the true potential by requiring that the predicted
density profile is consistent with  density profiles inferred from star
counts.

In the next subsection we explain how we selected candidate mass
distributions, and in Section~\ref{sec:DF} we present our multi-parameter \df.

\subsection{The mass model} \label{sec:mass_model}

It is obviously sensible to take the fullest possible advantage of existing
constraints on the Galaxy's mass distribution when selecting mass models.
Indeed, although the RAVE data dramatically tighten constraints on the
vertical distribution of matter, they are not well suited to constraining
the radial distribution of matter -- estimates of the circular speed and the
dynamics of halo tracers and satellite galaxies are much better suited to
that task. Here we use a methodology similar to that employed by
\citet{Caldwell1981}, \citet{Dehnen1998} and \citet{McMillan2011} to identify
a few-parameter family of mass models that can be confronted with the RAVE
data.

Our mass models have five components: a gas disc, a thin disc, a thick disc,
a flattened bulge and a dark halo. Since the stellar halo has negligible
mass, we do not explicitly include it in the mass model; its tiny mass is
negligible from a dynamical point of view and can be considered subsumed
within the dark halo. It is, however, important when we consider the phase
space distribution of stars and we therefore include it in our stellar
\df\ in Section~\ref{sec:DF}. For the density laws of the disc components we have
\begin{equation} \label{eq:rho_disc}
 \rho(R,z) = \frac{\Sigma_0}{2z_\mathrm{d}}
    \exp\left[-\left(\frac{R}{R_\mathrm{d}} + \frac{|z|}{z_\mathrm{d}} +
    \frac{R_\mathrm{hole}}{R}\right)\right],
\end{equation}
where $R$ and $z$ are the coordinates in a Galactocentric cylindrical
coordinate system and $\Sigma_0, R_\mathrm{d}, z_\mathrm{d}$ and
$R_\mathrm{hole}$ are parameters. A non-zero parameter $R_\mathrm{hole}$
creates a central cavity in the disc. This is used to model
the gas disc while for the other two discs it is set to zero. We further fix
the fraction of the local baryonic surface density contributed by the gas
disc to 25 per cent. The other fixed parameters for this component can be found in
Table~\ref{tab:fixed_model_params}.

The density distributions of  the dark halo and the bulge components are
\begin{equation} \label{eq:rho_spheres}
 \rho(R,z) = \frac{\rho_0}{m^\gamma(1+m)^{\beta-\gamma}}
 \exp[-(mr_0/r_\mathrm{cut})^2],
\end{equation}
where
\begin{equation}
 m(R,z) = \sqrt{(R/r_0)^2 + (z/qr_0)^2}.
\end{equation}
Here $\rho_0$ sets the density scale, $r_0$ is a scale radius, and the
parameter $q$ is the axis ratio of the isodensity surfaces. The exponents
$\gamma$ and $\beta$ control the inner and outer radial slopes of the radial
density profile.

For our standard model we adopt a spherical NFW model \citep{NFW1996} ($q=1, 
\alpha=1,\beta=3,r_{\rm cut}\simeq\infty$) for the dark halo. In
Section~\ref{sec:flattening-independent} we also consider flattened halo
configurations.

The bulge model is not varied in the course of our model fitting process,
because we use no data that might constrain its parameters. Following
\citet{McMillan2011} we use a model similar to that constructed by
\citet{Bissantz2002}. It has an axis ratio $q=0.5$ and extends to
$r_{\rm cut}=2.1\kpc$: Table~\ref{tab:fixed_model_params} lists the other
parameters of the bulge.

\begin{table}
 \centering
 \caption{Parameters that are fixed in our Galaxy mass model.}
 \label{tab:fixed_model_params}
 \begin{tabular}{lc}
  \hline
  Gas disc\\[.1cm]
  $\Sigma_\mathrm{g}(R_0)$			 	& $(\Sigma_{\rm thin}(R_0) + \Sigma_{\rm thick}(R_0))/3$ \\
  $R_\mathrm{d,g}$					& $2R_\mathrm{d,thin}$ \\
  $z_\mathrm{d,g}$ [kpc]				& 0.04 \\
  $R_\mathrm{hole,g}$ [kpc]				& 4 \\[.2cm]
  Bulge\\[.1cm]
  $\rho_\mathrm{0,b}$ [M$_\odot$ kpc$^{-3}$]	 	& $9.49 \times 10^{10}$ \\
  $q_\mathrm{b}$					& 0.5 \\
  $\gamma_\mathrm{b}$					& 0 \\
  $\beta_\mathrm{b}$					& 1.8 \\
  $r_\mathrm{0,b}$ [kpc]				& 0.075 \\
  $r_\mathrm{cut,b}$ [kpc]				& 2.1 \\[.2cm]
  Dark halo\\[.1cm]
  $q_\mathrm{dm}$					& 1 \\
  $\gamma_\mathrm{dm}$					& 1 \\
  $\beta_\mathrm{dm}$					& 3 \\
  $r_\mathrm{cut,dm}$ [kpc]				& $10^5$ \\
  \hline
 \end{tabular}
\end{table}

\begin{figure}
 \centering
 \includegraphics[width=0.47\textwidth]{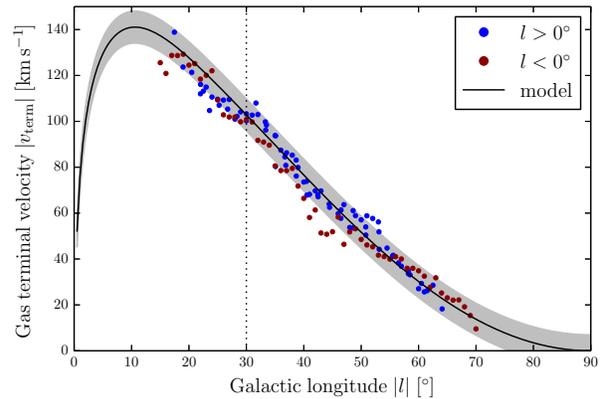}
 \caption{Comparison of the measured terminal velocities and the prediction
 by one of our mass models (solid black line). The grey shaded area
 illustrates the velocity dispersion of $7\kms$ that we assumed. Measurements
 left of the dotted vertical line were not included for the fit. This
 particular model has the parameters given in Table~\ref{tab:best-fit-params},
 but the same quality of the fit is achieved for all reasonable choices for
 $\rhodm$ and the parameters for the vertical structure of the disc.}
 \label{fig:Terminal_velocities}
\end{figure}

\begin{figure*}
 \centering
 \includegraphics[width=0.98\textwidth]{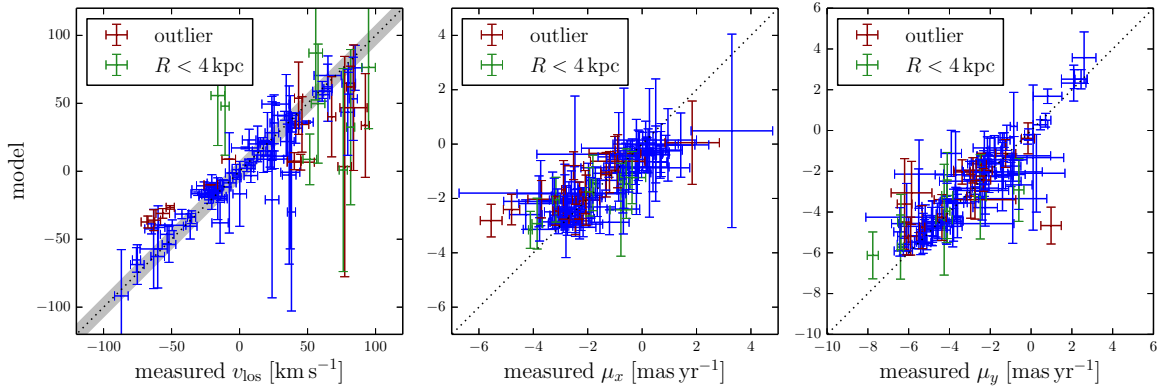}
 \caption{Comparison of the maser measurements by \citet{Reid2014} with our
 model predictions. The vertical error bars show the uncertainties arising from
 the uncertainties in the parallax estimations. In the left-most panel the
 grey shaded area illustrates the velocity dispersion of $7\kms$ that we
 assumed. In the middle and right-most panels this velocity dispersion is
 already included in the vertical error bars because for proper motions this
 uncertainty is dependent on the individual distances of the
 masers. The sources plotted in red were considered outliers by
 \citet{Reid2014} and
 those closer than 4 kpc to the Galactic centre were excluded from the fit.}
 \label{fig:Masers}
\end{figure*}

Since the thin and the thick disc density laws have three free parameters
each ($\Sigma_0,R_{\rm d},z_{\rm d}$) and the dark halo has two free
parameters ($\rho_0,r_0$), our mass model has eight parameters. Since our
constraints are not suited to fixing differing scale radii for the two
discs, we reduce the number of free parameters to seven by setting
$R_\mathrm{d,thick} = R_\mathrm{d,thin}$. We then use the constraints from
the literature listed in Section~\ref{sec:Data} to fix all remaining
parameters except the local dark matter density, $\rhodm$, the scale heights
$\zthin$ and $\zthick$ of the thin and thick discs, and the parameter
$\fthick$, which determines the fraction $(1+\fthick^{-1})^{-1}$ of the local
stellar mass density that is contributed by the thick disc. The approach
closely follows that of \citet{McMillan2011} except that we use the \amoeba\
algorithm \citep{NumericalRecipes} to find the parameter set with maximum
likelihood instead of evaluating the full posterior via MCMC as in
\citet{McMillan2011}. Figures~\ref{fig:Terminal_velocities} and
\ref{fig:Masers} illustrate a typical quality of fit that we achieve for the
terminal velocities and the maser data.

The three parameters, $\zthin,\zthick$ and $\fthick$ are finally fixed using
the RAVE data and the stellar \df, which we describe in the next subsection.
\subsection{DF for the discs}\label{sec:DF}

Following \citet{Binney2010} we assume that the \df s of the discs can be
well approximated by analytic functions of the three action integrals $J_i$.
We use the ``St\"ackel Fudge'' introduced by \citet{Binney2012a} to evaluate
the $J_i$ given phase-space coordinates $(\vx,\vv)$. (Details of some
technical improvements are given in \citet{Binney2014c}.)

Our \df s are built up out of ``quasi-isothermal'' components. The \df\ of
such a component is
\[\label{eq:qi}
 f(J_r,J_z,L_z)=f_{\sigma_r}(J_r,L_z)f_{\sigma_z}(J_z,L_z),
\] 
where $f_{\sigma_r}$ and $f_{\sigma_z}$ are defined to be
\[\label{planeDF}
 f_{\sigma_r}(J_r,L_z)\equiv \frac{\Omega\Sigma}{\pi\sigma_r^2\kappa}
 [1+\tanh(L_z/L_0)]\e^{-\kappa J_r/\sigma_r^2}
\]
and
\[\label{basicvert}
 f_{\sigma_z}(J_z,L_z)\equiv\frac{\nu}{2\pi\sigma_z^2}\,
 \e^{-\nu J_z/\sigma_z^2}.
\]
Here $\Omega(L_z)$, $\kappa(L_z)$ and $\nu(L_z)$ are, respectively, the
circular, radial and vertical epicycle frequencies of the circular orbit with
angular momentum $L_z$, while
\[\label{eq:defsSigma}
 \Sigma(L_z)=\Sigma_0\e^{-\Rc/\Rd},
\]
where $\Rc(L_z)$ is the radius of the circular orbit, determines the surface
density of the disc: to a moderate approximation this surface density can be
obtained by using for $L_z$ in equation (\ref{eq:defsSigma}) the angular
momentum $L_z(R)$ of the circular orbit with radius $R$. The functions
$\sigma_r(L_z)$ and $\sigma_z(L_z)$ control the radial and vertical velocity
dispersions in the disc and are approximately equal to them at $\Rc$. Given
that the scale heights of galactic discs do not vary strongly with radius
\citep{vanderKruit1981}, these quantities must increase inwards. We adopt the
dependence on $L_z$
\begin{eqnarray}
 \sigma_r(L_z)&=&\sigma_{r0}\,\e^{(R_0-\Rc)/R_{\sigma,r}}\cr
 \sigma_z(L_z)&=&\sigma_{z0}\,\e^{(R_0-\Rc)/R_{\sigma,z}},
\end{eqnarray}
so the radial scale-lengths on which the velocity dispersions
decline are $R_{\sigma,i}$. Our expectation is that $R_{\sigma,i} \sim 2\Rd$.

In equation (\ref{planeDF}) the factor containing tanh serves to eliminate
retrograde stars; the value of $L_0$ controls the radius within which
significant numbers of retrograde stars are found, and should be no larger
than the circular angular momentum at the half-light radius of the bulge.
Provided this condition is satisfied, the results for the extended solar
neighbourhood presented here are essentially independent of $L_0$.

We take the \df\ of the thick disc to be a single pseudo-isothermal. The thin
disc is treated as a superposition of the cohorts of stars that have age
$\tau$ for ages that vary from zero up to the age $\tau_{\rm max} \simeq
10\Gyr$ of the thin disc. We take the \df\ of each such cohort to be a
pseudo-isothermal with velocity-dispersion parameters $\sigma_{r}$ and
$\sigma_{z}$ that depend on age as well as on $L_z$.  Specifically, from
\citet{Aumer2009} we adopt
\begin{eqnarray}\label{sigofLtau}
 \sigma_r(L_z,\tau)&=&\sigma_{r0}\left(\frac{\tau+\tau_1}{\tau_{\rm m}+\tau_1}\right)^\beta\e^{(R_0-\Rc)/R_{\sigma,r}}\nonumber\\
 \sigma_z(L_z,\tau)&=&\sigma_{z0}\left(\frac{\tau+\tau_1}{\tau_{\rm m}+\tau_1}\right)^\beta\e^{(R_0-\Rc)/R_{\sigma,z}}.
\end{eqnarray}
Here $\sigma_{z0}$ is the approximate vertical velocity dispersion of local
stars at age $\tau_{\rm m}\simeq10\Gyr$, $\tau_1$ sets velocity dispersion at
birth, and $\beta\simeq0.33$ is an index that determines how the velocity
dispersions grow with age. We further assume that the star-formation rate in
the thin disc has decreased exponentially with time, with characteristic time
scale $t_0$, so the thin-disc \df\ is
\[\label{thinDF}
 f_{\rm thn}(J_r,J_z,L_z)=\frac{\int_0^{\tau_{\rm m}}\rd\tau\,\e^{\tau/t_0}
 f_{\sigma_r}(J_r,L_z)f_{\sigma_z}(J_z,L_z)}{t_0(\e^{\tau_{\rm m}/t_0}-1)},
\]
where $\sigma_r$ and $\sigma_z$ depend on $L_z$ and $\tau$ through equation
(\ref{sigofLtau}). We set the normalising constant $\Sigma_0$ that appears in
equation (\ref{eq:defsSigma}) to be the same for both discs and use for the
complete \df
\[
 f_{\rm disc}(J_r,J_z,L_z)=f_{\rm thn}(J_r,J_z,L_z) +
 F_{\rm thk}f_{\rm thk}(J_r,J_z,L_z),
\]
where $F_{\rm thk}$ is a parameter that controls the fraction $(1+F_{\rm
thk}^{-1})^{-1}$ of stars that belong to the thick disc.\footnote{Note, that $F_{\rm thk}$
is the ratio of the total masses of the thick and the thin discs, while the
parameter $\fthick$ used for the mass model is the ratio of the local mass
densities of the two discs. Hence the two parameters are intimately related
but not the same.}

The \df s of the thin and thick discs each involve five important parameters,
$\sigma_{r0}$, $\sigma_{z0}$, $\Rd$, $R_{\sigma,r}$ and $R_{\sigma,z}$. The
\df\ of the thin disc involves four further parameters, $\tau_1$,
$\tau_{\rm m}$, $\beta$ and $t_0$, but we shall not explore the impact of
changing these here because we do not consider data that permit
discrimination between stars of different ages. Therefore following
\citet{Aumer2009} we adopt throughout $\tau_1=0.01\Gyr$,
$\tau_{\rm m}=10\Gyr$, $\beta=0.33$ and $t_0=8\Gyr$.

\subsection{DF of the stellar halo}

Due to the magnitude limits of RAVE, most of its stars belong to the thin and
thick discs. The sample does, however, contain a small but non-negligible
population of halo stars, which are identifiable by their low or even
negative values of the azimuthal velocity $V_\phi$ \citep{Piffl2014}. We
have added to the \df\ a component for the stellar halo to prevent the
fitting routine distorting the thick disc in an attempt to account for the
presence in the sample of halo stars.

The density of the stellar halo is generally thought to follow a power-law in
Galactocentric radius, i.e.\ $\rho_{\rm halo} \propto r^{-\alpha}$, with the
power-law index $\alpha \simeq 3.5$ \citep[e.g.][\S10.5.2]{GalacticAstronomy}.
We can model such a configuration using the following form of the \df\
(Posti et al., in preparation)
\[\label{eq:haloDF1}
 f_{\rm halo}(J_r,J_z,L_z) = \left(\frac{h(L_{\rm c0})}{
 h(\vJ)}\right)^{\alpha},
\]
where $h(\vJ)$ is a homogeneous function of degree one [i.e.\ $h(\beta\vJ)
= \beta h(\vJ)$], and $L_{\rm c0}$ is the angular momentum of a circular orbit
of radius $R_0$. When the circular speed is independent of radius, the
density generated by the {\df}~(\ref{eq:haloDF1}) declines with radius as
$r^{-\alpha}$, so we adopt $\alpha=3.5$.
Our choice
\[\label{eq:haloDF2}
 h(\vJ) = J_r + \frac{\Omega_z}{\Omega_r} J_z + \frac{\Omega_\phi}{\Omega_r}
 |L_z|,
\]
where $\Omega_i(\vJ)$ are the characteristic frequencies of the orbit with
actions $\vJ$, ensures that in a spherical potential the halo would be
approximately spherical. Since the Galaxy's potential is somewhat flattened,
our halo will be slightly flattened too.  The RAVE data alone are not well
suited to constraining the stellar halo, so we defer this exercise to a later
paper. We include the stellar halo only in order to prevent distortion of the
thick disc that is fitted to the data. Our complete total \df\ is
\begin{equation}
 f(J_r,J_z,L_z) = f_{\rm disc}(J_r,J_z,L_z) + F_{\rm halo}f_{\rm halo}(J_r,J_z,L_z)
\end{equation}
\subsection{Model-RAVE comparison} \label{sec:Model-RAVE-comparison}
\begin{figure*}
 \centering
 \includegraphics[width=0.98\textwidth]{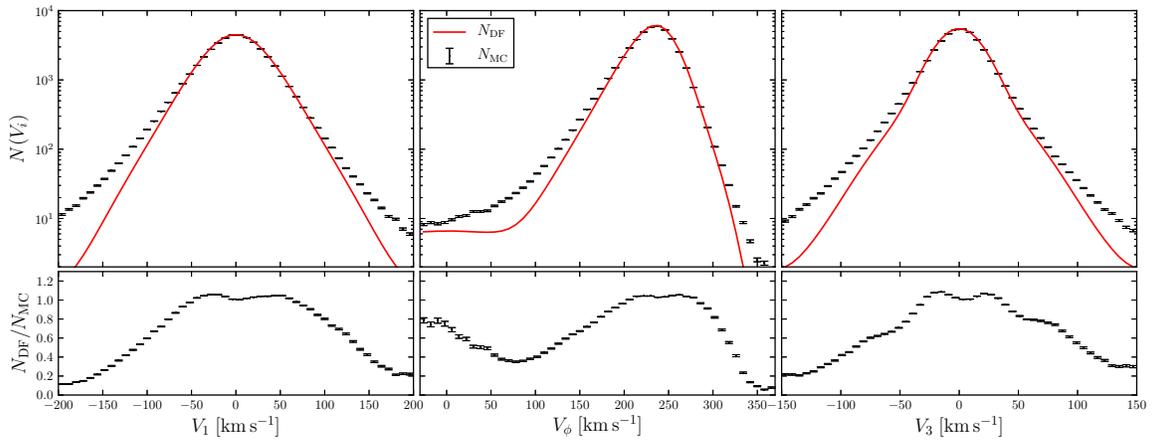}
  \caption{Distortion on the velocity distributions by uncertainties in
the RAVE distances. The three upper panels are histograms of the three velocity
components for stars with $R<R_0$ and $|z|<0.3\kpc$. The
solid red lines show the number of stars predicted by the \df\ at the
barycentre of the bin, and the error bars show the numbers of stars selected
by Monte Carlo re-sampling. For the latter we take the mean of
100 realisations. The lower panels show the value of the \df\ divided by the
number of stars in the Monte-Carlo sample.}
 \label{fig:Compare_NDF_NMC}
\end{figure*}

RAVE, like any spectroscopic survey, has a non-trivial selection function:
potential targets were divided into bands by apparent magnitude and then the
spectrograph's fibres were allocated to as many stars as possible in a given
band with the exposure time being band-specific. From the resulting spectra
the pipeline extracts acceptable stellar parameters with a probability that
to some extent depends on metallicity. Hence the probability that a given
star enters the final catalogue -- the selection function -- depends on the
star's apparent magnitude, location and metallicity. These probabilities will
be derived and discussed in a forthcoming paper (Piffl et al., in
preparation), and we proceed here without using these probabilities.

Of the above-mentioned selection criteria, only the
metallicity is -- via the birth time and place of a star -- correlated with
the stellar velocities. Fortunately, the great majority of spectra yield
stellar parameters, so the dependence of the selection function on
metallicity is weak. If we neglect this weak dependence, the selection
function is independent of stellar velocity, so we can predict the velocity
distribution $n(\vv)=f(\vx,\vv)/\rho(\vx)$ of the catalogued stars from a
model's \df, $f(\vx,\vv)$. It is interesting to ask to what extent the \df\
of the RAVE stars is constrained by the velocity distributions presented by
\citet{Binney2014b}. We focus on the velocity distributions of the giant
stars ($\log\,g<3.5$) which provide wider spatial coverage than the dwarf
stars. In Section~\ref{sec:kinematics} we will compare the RAVE data for hot
dwarf stars ($\logg > 3.5$ and $\Teff > 6000\K$) with the predictions made by
the most successful of the \df s we obtain by fitting the giants.

We define eight spatial bins in the $(R,z)$ plane. Four bins for stars inside
the solar cylinder with $R_0-1\kpc < R < R_0$ and $|z|$ in
[0,0.3],[0.3,0.6],[0.6,1.0] or [1,1.5] kpc. The other four bins cover the
same $z$ ranges but cover the regions 1 kpc outside the solar cylinder, i.e.\
$R_0 < R < R_0 + 1\kpc$. After sorting the stars into these bins, we compute
the velocity distributions predicted by the \df\ at the mean $(R,z)$
positions (barycentre) of the stars in each bin. For each bin we have a
histogram for each component of velocity, so we accumulate $\chi^2$ from 24
histograms.  Throughout this work we compute velocities in the coordinate
system that \citet{Binney2014b} found to be closely aligned with the velocity
ellipsoid throughout the extended solar neighbourhood -- this system is quite
closely aligned with spherical coordinates. We denote the velocity component
along the long axis of the velocity ellipsoid -- pointing more or less
towards the Galactic centre -- with $V_1$, the azimuthal component with
$V_\phi$, and the remaining component with $V_3$ which is close to the latitudinal direction \citep[cf.\ also][]{Bond2010}.

The resulting model distributions cannot be directly compared to the
observed distributions, because the latter are widened by errors in the
velocity and parallax estimates. We fold the model distributions with the
average velocity uncertainties of the bin's stars to obtain $N_{\rm
bary}(V_i)$. The distortions arising from the parallax error are less
straight forward to introduce: following \citet{Binney2014b} we create a
Monte Carlo realisation of a given \df\ by randomly assigning to each star in
our RAVE sample a new ``true'' distance according to its (sometimes
multi-modal) distance pdf, and a new ``true'' velocity according to the model
velocity distribution at this position. With these new phase space
coordinates we compute new observed line-of-sight velocities and proper
motions. These are finally equipped with random observational errors. Using
the original catalogue distances, we then compute new realistically
distorted velocity distributions, $N_{\rm MC}(V_i)$, based on the \df\ that
can be compared directly to the original RAVE distributions in a number of
spatial bins. We minimise the Poisson noise in $N_{\rm MC}(V_i)$ by choosing
100 new velocities for each star. This procedure is computationally expensive
and the distortions vary only weakly for reasonable choices of the \df\
parameters. To speed up the process, we store the ratio $N_{\rm
bary}(V_i)/N_{\rm MC}(V_i)$ for a \df\ that is already a good match of the
RAVE data. Examples of these ratios are shown in the lower panels of
\figref{fig:Compare_NDF_NMC} while the upper panels plot the actual
distributions. The ratio is near unity in the core of the distribution but
falls to $<0.2$ in the wings because distance errors scatter stars to high
apparent velocities. These ratios are then used to correct all \df\
predictions before they are compared with the data.

This approach is far from perfect, because by comparing velocity histograms
instead of assigning likelihoods to individual stars we loose the
information encoded in the correlations between the velocity components.
However, an approach based on computing likelihoods for the full
phase-space distribution is currently computationally too expensive to
allow for testing a large number of models.

\begin{figure}
 \centering
 \includegraphics[width=0.47\textwidth]{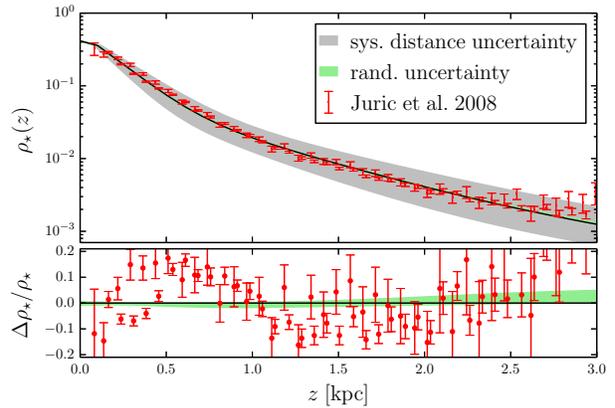}
 \caption{Vertical stellar density profile as predicted by the \df\ (upper
 panel) and
 relative random residuals (lower panel). The green  shaded area shows the random
 variations introduced by the uncertainty of the \df\ parameters when fitted to
 the RAVE data. The measurements and uncertainties from \citet{Juric2008} are also
 shown for comparison. The grey shaded area illustrates the uncertainty introduced
 by the uncertain distance scale of \citet{Juric2008}. Note, that the model profiles
 were only constrained by stars with $|z|<1.5\kpc$.}
 \label{fig:Profile_variation}
\end{figure}

With our approach and the RAVE giant sample we can determine the values
of the \df\ parameters to very high precision. Using Markov Chain Monte Carlo
re-sampling (MCMC) we find that the pseudo-velocity dispersions are fixed to a
fraction of a percent, while the thick disc dispersion scale lengths and $F_{\rm
thk}$ are fixed to $\sim 2$ per cent. Only the stellar halo normalisation has a relatively
large uncertainty of $\lesssim7$ per cent as is to be expected from the small number of
halo stars in RAVE. \figref{fig:Profile_variation} illustrates the resulting
uncertainties on the vertical disc profile. Clearly the uncertainties on the
\citetJuric\ distance scale dominate the error budget by far.

\subsection{Finding a pair}

We now describe how we find what we call a ``self-consistent'' mass
model--\df\ pair that has a pre-defined value of the dark-matter density at
the position of the Sun, $\rhodm$, and is consistent with the kinematics the
RAVE giants. With the ingredients outlined above we cannot construct a truly
self-consistent model in the sense that both the collision-less Boltzmann
equation and the Poisson equation are fulfilled as in \cite{Binney2014c},
because we do not include the dark halo and the bulge in our \df. We call a
mass model--\df\ pair ``self-consistent'' if the mass distribution of the
stellar disc implied by the \df\ is consistent with the mass distribution of
the stellar disc assumed in the mass model. 

In this spirit we identify the disc scale lengths in the \df\ with the scale
lengths in the mass model, and we equate in the \df\ and the mass model the
parameters $F_{\rm thk}$ that determine the fraction of stars that belong to
the thick disc. 

It is a priori to be expected that the RAVE data do not strongly constrain
the values of the scale lengths $R_{\sigma,r}$ and $R_{\sigma,z}$ of the
velocity dispersions in the thin disc because, by virtue of the survey's
avoidance of regions of low Galactic latitude $b$, radii $R$ that differ
materially from $R_0$ are only probed at high $|z|$, so radial and vertical
gradients of the dispersions are hard to disentangle, especially as the thick
disc dominates at high $|z|$.  Experiments with the data confirm that the
$R_{\sigma,i}$ of the thin disc are poorly constrained. In fact, if allowed
to vary, their values tend to infinity, implying that the thin disc flares
strongly, contrary to observation. We decided to fix the value of the
$R_{\sigma,i,\rm thn}$ to $9\kpc$ which is 3--4 times the values
of the scale length the data imply for the mass model. We tested smaller and
larger values of $R_{\sigma,i}$ for the thin disc and found no significant
influence on our results.

\begin{figure}
 \centering
 \includegraphics[width=0.47\textwidth]{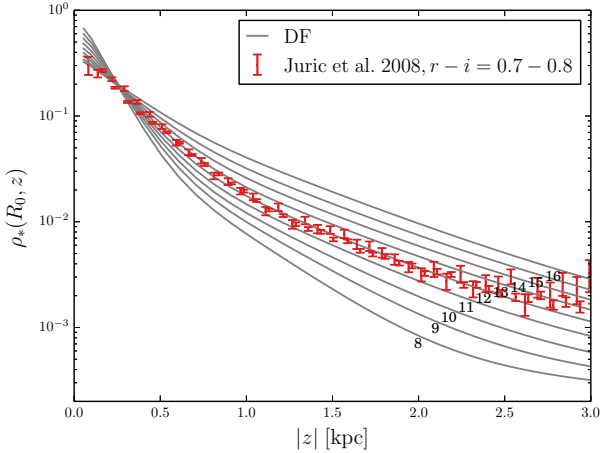}
 \caption{The full curves show the variation with $|z|$ of the number density
 of stars predicted by the best-fitting \df\ in mass models with various
assumed values of $\rhodm$; each curve is labelled by $\rhodm$
 in units of $10^{-3}\msun\pc^{-3}$ and it can be shifted up or down
 at will. The red error bars show the number density of stars measured by
\citet{Juric2008}.}
 \label{fig:all_zprofs1}
\end{figure}

With these assumptions our Galaxy model contains 4+7 free
parameters: $\rhodm\equiv\rho_{\rm dm}(R_0,z_0)$, $z_{\rm d,thn}$,
$z_{\rm d,thk}$ and $F_{\rm thk}$ for the mass model and $\sigma_{r0,\rm thn}$,
$\sigma_{z0,\rm thn}$, $\sigma_{r0,\rm thk}$, $\sigma_{z0,\rm thk}$,
$R_{\sigma,r\rm thk}$, $R_{\sigma,z\rm thk}$ and $F_{\rm halo}$ for the \df.

Given a mass model, it is computationally relatively cheap to adjust the
parameters in the \df\ to optimise the fit between the predicted and observed
velocity histograms. By contrast, any change in the mass model requires the
relatively costly computation of new actions at a large number of points in
phase space. Hence we proceed as follows: for a trial mass model, we use
\amoeba\ minimisation to choose the \df\ that provides the best fit to the
observed velocity histograms. Next, holding constant $\rhodm$, we apply
\amoeba\ to adjust the mass model to optimise the fit between the vertical
density profile of the stars predicted by the \df\ and assumed by the mass
model. In this process we keep the velocity-dispersion parameters of the \df\
fixed, with the result that the fit between the predicted and RAVE kinematics
deteriorates, but fortunately only moderately even when the predicted stellar
profile is materially altered.  Once the stellar density profiles associated
with the mass model and the \df\ have been brought to good agreement, the
parameters of the \df\ are fine-tuned by another run of \amoeba\ to re-optimise
the fit between the predicted and observed kinematics on the spatially binned
data, and then the mass model is readjusted to restore optimum agreement
between the vertical density profiles.

The outcome of this procedure is a \df\ and a mass model that are consistent
with one another as regards the spatial distribution of stars, and consistent
with the observed kinematics of the RAVE stars. As $\rhodm$ is increased, the
mass of the disc decreases to ensure that the constraints from the terminal
velocities and proper motion of Sgr A* continue to be satisfied, and the
vertical density profile of the model discs becomes steadily shallower. For a
small range of values of $\rhodm$, the model's profile is consistent with the
star counts. Fig.~\ref{fig:all_zprofs1} shows this process in action: the
black curves show the density profile predicted by the \df\ of the
self-consistent mass model--\df\ pair found for the value of $\rhodm$ that is
indicated by the numbers $8,9,\ldots$ on each curve, where the units are
$10^{-3}\msun\pc^{-3}$.

\section{Results}

\begin{figure}
 \centering
 \includegraphics[width=0.47\textwidth]{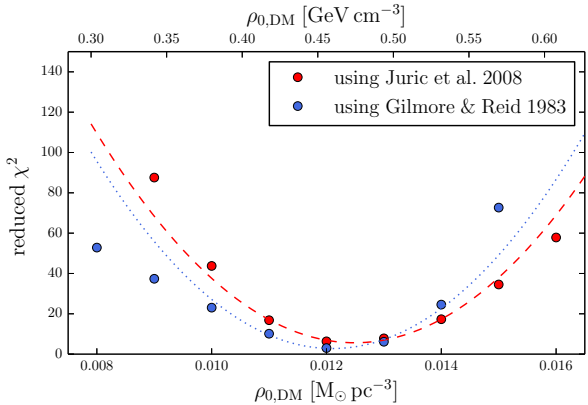}
 \caption{Red dots: Reduced $\chi^2$ distance between the vertical stellar mass profile
 predicted by the \df\ and the observational profiles by \citet{Juric2008}
 as a function of the local density of a spherical dark-matter halo. Green
 dots show the reduced $\chi^2$ distance from the density profile of
\citet{Gilmore1983}. The red and green dashed lines are parabolas fitted to
the red/green dots.}
 \label{fig:chi2_plot}
\end{figure}

The red points in \figref{fig:chi2_plot} show $\chi^2$ for the fit provided
by each mass model--\df\ pair to the \citetJuric\ data; the red dashed curve
shows the best-fit parabola through these points. Its minimum lies at
$\rhodm=0.01262\msun\pc^{-3}$ and its curvature implies a remarkably small
uncertainty, $\sim 0.4$ per cent. This is just the statistical uncertainty from the
comparison to the observational profiles and should \emph{not} be used,
because -- as we will show below -- the systematic uncertainties are much
larger.

\begin{table}
 \centering
 \caption{Best-fit parameters.}
 \label{tab:best-fit-params}
 \begin{tabular}{ccl}
  \hline
  Model potential parameters \\
  \hline
  $\Sigma_{\rm 0,thin}$ & 570.7 & $\msun\pc^{-2}$ \\ 
$\Sigma_{\rm 0,thick}$ & 251.0 & $\msun\pc^{-2}$ \\ 
$R_{\rm d}$ & 2.68 & $\kpc$ \\ 
$z_{\rm d,thin}$ & 0.20 & $\kpc$ \\ 
$z_{\rm d,thick}$ & 0.70 & $\kpc$ \\ 
$\Sigma_{\rm 0,gas}$ & 94.5 & $\msun\pc^{-2}$ \\ 
$R_{\rm d,gas}$ & 5.36 & $\kpc$ \\ 
$\rho_{\rm 0,dm}$ & 0.01816 & $\msun\pc^{-3}$ \\ 
$r_{\rm 0,dm}$ & 14.4 & $\kpc$ \\ 

  \hline
  \df\ parameters \\
  \hline
  $\sigma_{r,\rm thin}$ & 34.0 & $\kms$ \\ 
$\sigma_{z,\rm thin}$ & 25.1 & $\kms$ \\ 
$R_{\sigma,r,\rm thin}$ & 9.0 & $\kpc$ \\ 
$R_{\sigma,z,\rm thin}$ & 9.0 & $\kpc$ \\ 
$\sigma_{r,\rm thick}$ & 50.6 & $\kms$ \\ 
$\sigma_{z,\rm thick}$ & 49.1 & $\kms$ \\ 
$R_{\sigma,r,\rm thick}$ & 13.0 & $\kpc$ \\ 
$R_{\sigma,z,\rm thick}$ & 4.2 & $\kpc$ \\ 
$F_{\rm thick}$ & 0.447 \\ 
$F_{\rm halo}$ & 0.026 \\ 

  \hline
 \end{tabular}
\end{table}
\begin{figure*}
 \centering
 \includegraphics[width=0.98\textwidth]{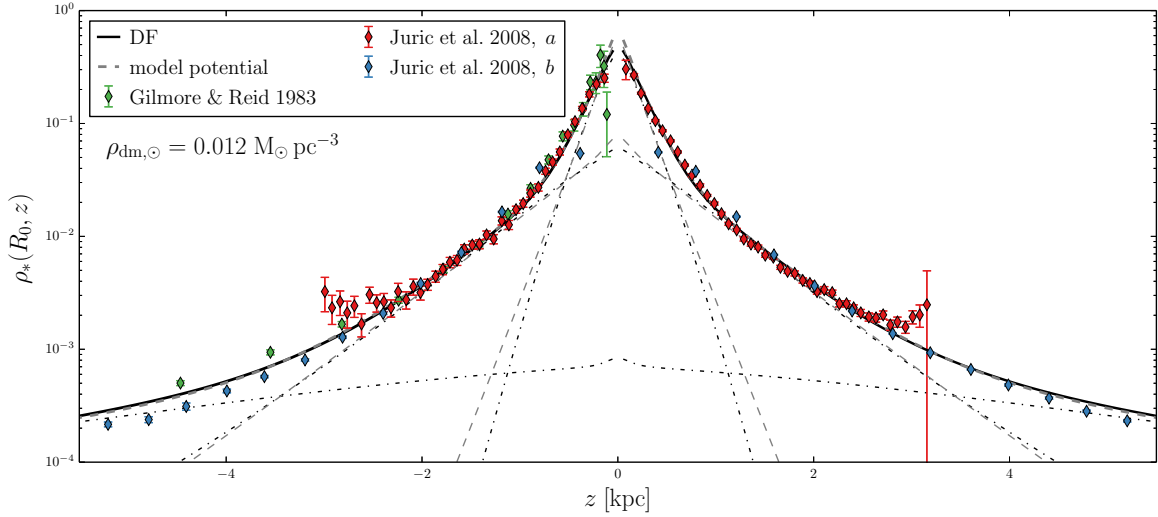}
 \caption{The full black curve shows the vertical density profile of the disc
 predicted by the \df\ for $\rhodm = 0.012\msun\pc^{-3}$; the mostly overlying
 dashed black curve shows the corresponding density profile in the mass
 model. The other dashed black lines show the profiles of the thin and thick
 discs in the mass model. The dotted curves show the corresponding predictions
 of the \df\ for both discs and the stellar halo (which has no explicit
 counterpart in the mass model). The red and blue error bars show the vertical profile
 measured by \citet{Juric2008} for stars with $r-i \in [0.7,0.8]$ (``a'',
 red symbols) and with $r-i \in [0.15,0.2]$ (``b'', blue symbols). The latter wasn't used
 in the analysis and is show only for illustrative purposes. The green error bars
 show the profile measured by \citet{Gilmore1983}.}
 \label{fig:all_zprofs2}
\end{figure*}

Fig.~\ref{fig:all_zprofs2} shows as a heavy black line the stellar density
profile provided by the mass model--\df\ pair for local dark-matter density
$0.012\msun\pc^{-3}$. The fit to the red data points, which show the
star-count data from \citetJuric, is excellent both below and above the
Galactic plane. The dashed grey lines in \figref{fig:all_zprofs2} show the
densities contributed by the thin and thick stellar discs of the mass model,
while the dotted black curves show the densities yielded by the \df\ for the
thin and thick discs and the stellar halo. At $z=0$ the dashed curves from
the mass model are unrealistically cusped on account of our assumption of
naive double-exponential discs. Otherwise the agreement between the densities
provided for the thick disc between the mass model and the \df\ is perfect.
The agreement between the curves for the thin disc is nearly perfect within
$\sim1.5$ scale heights of the plane, but at greater heights, where the thick
disc strongly dominates, the \df\ provides slightly lower density than the
mass model. This discrepancy implies that the \df\ breaks the total stellar
profile into thin- and thick-disc contributions in a slightly different way
to the mass model. Since a real physical distinction between these components
can only be made on the basis of age or chemistry
\citep[e.g.][]{GalacticAstronomy}, the minor difference between the two
thin-disc curves in \figref{fig:all_zprofs2} should not be considered
significant at this stage.

The green error bars in \figref{fig:all_zprofs2} show the stellar densities
inferred by \citet{Gilmore1983} for stars with absolute visual magnitude
$M_V$ between 4 and 5 with an assumed vertical metallicity gradient of
$-0.3\dex/\kpc^{-1}$ (in their Table~2). The green dots in
\figref{fig:chi2_plot} show the $\chi^2$ values we obtain when we adopt the
Gilmore--Reid data points. They indicate a deeper minimum in $\chi^2$
occurring at a smaller dark-halo density: $\rhodm=0.01200\msun\pc^{-3}$.

\subsection{Systematic uncertainties} \label{sec:systematics}
\begin{figure}
 \centering
 \includegraphics[width=0.47\textwidth]{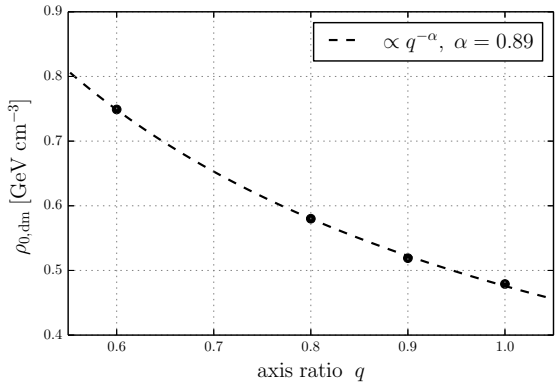}
 \caption{Best-fitting value for the local dark-matter density $\rhodm$ as a
 function of the assumed axis-ratio $q$ of the dark-matter halo. A value of
 $q=1$ implies a spherical halo, while smaller values lead to oblate
 configurations. The dashed black line shows a power-law fitted using least square minimisation.}
 \label{fig:halo_flattening}
\end{figure}
The results presented above are based on a very sophisticated model that
involves a number of assumptions and approximations. Deviations of the truth
from these assumptions and approximations will introduce systematic errors
into our results. We can assess the size of such systematic errors much more
easily in some cases than in others. We have not assessed the errors arising
from:
\begin{itemize}
 \item the functional form of the mass model;
 \item the functional form of the \df;
 \item the age-velocity dispersion relation in the thin disc;
 \item the adopted value of $L_0$ in disc \df: variation will affect the
 normalisation of stellar halo;
 \item the power-law slope and quasi-isotropy of the stellar halo -- we will
 investigate this in a future paper;
 \item the solar motion w.r.t.\ the LSR.
\end{itemize}
 The first two points include the approximation that the Galactic disc is
smooth and axis-symmetric, so the flows induced by
spiral waves that have been detected by RAVE \citep{Siebert2012, Williams2013} and in
simulations \citep{Faure2014, Debattista2014} are neglected.

We have investigated the sensitivity of our results to:

\begin{itemize}
 \item $R_0$, which controls the circular speed: a value of $R_0=8\kpc$
reduces $\rhodm$ by 10 per cent.

 \item The contribution of the gas disc to the local baryonic surface
density. If we assume 33 per cent instead of our standard value of
25 per cent, we find slightly different structural parameters for the stellar discs,
but our best-fit value for $\rhodm$ remains unchanged.

 \item $R_{\sigma,i}$ for the thin disc: using $R_{\sigma,i}=6\kpc$ reduces $\rhodm$ by $<2$ per cent.

 \item The fact that $r_{0,\rm dm}$ changes with $\rhodm$ on account of the
halo constraints:  setting $r_{0,\rm dm}=20\kpc$ increases $\rhodm$
by 2 per cent.

 \item Equal scale radii for thin and thick disc: setting $R_{\rm
d,thick}/R_{\rm d,thin} = 0.6$ (resulting in $R_{\rm d,thick} \simeq
2\kpc$ and $R_{\rm d,thin}\simeq 3.5\kpc$ similar to \citet{Bovy2012c}),
increases $\rhodm$ by 4 per cent.

 \item Flattening the dark halo: a flatter dark halo increases $\rhodm$
significantly. See \figref{fig:halo_flattening}. 

 \item Systematic uncertainties in the distance scale of \citetJuric: if this
distance scale is increased by a factor $\alpha$, $\rhodm$ proves to be
almost proportional to $\alpha$, with a 20 per cent increase in $\alpha$ causing
$\rhodm$ to increase by 8 per cent. A different value for the binary fraction
has a very similar, but smaller, effect to a general change of the distance
scale, and is hence also covered in this uncertainty.
For a flattened dark matter halo with $q=0.6$ the same test yielded a variation of 10 per cent.
\end{itemize}

The three most critical systematic uncertainties are therefore the axis ratio
$q$ of the dark halo, the solar distance to the Galactic centre and the distance scale used to construct the
observational vertical stellar density profile. Simply adding in quadrature
the uncertainties other than halo flattening listed above leads to a combined
systematic uncertainty of $\sim 15$ per cent. Combining this with the uncertainty
associated with dark-halo flattening we arrive at our result
\[
 \rhodm=
\begin{cases}
 (0.48 \times q^{-\alpha})~{\rm GeV\,cm^{-3}} \pm 15\%&\\
(0.0126 \times q^{-\alpha})~{\rm M_\odot\,pc^{-3}} \pm 15\%
\end{cases}
\]
with $\alpha = 0.89$ and $q$ the axis ratio of the dark halo. We remark that the given uncertainties are not statistically robust confidence intervals, but are estimates based on the above test runs.
Because our kinematic data are resticted to $|z|<1.5\kpc$ it is not possible for us to reliably single out a preferred value of $q$. For this, data at larger distances from the Galactic plane would be needed (cf. \figref{fig:EllipsoidalMass}). In Section~\ref{sec:Discussion} we discuss this further.

Note, there is an additional potential source of uncertainty that we have
not included in our estimate: \citet{Schoenrich2014b} find hints that the
common practice of assuming uncorrelated errors in the stellar parameters
when deriving distance estimates  leads
to over-confident results. Hence the parallax uncertainties reported by
\citet{Binney2014a} might be under-estimates. To test the possible influence
we doubled the individual parallax uncertainties (a worst case scenario) and
repeated the fit. The best-fitting value for $\rhodm$ increased by $\sim7$ per cent.
A similar uncertainty is shared by all studies that use distances inferred
from stellar parameters.
\subsection{Flattening-independent results} \label{sec:flattening-independent}
\begin{figure}
 \centering
 \includegraphics[width=0.47\textwidth]{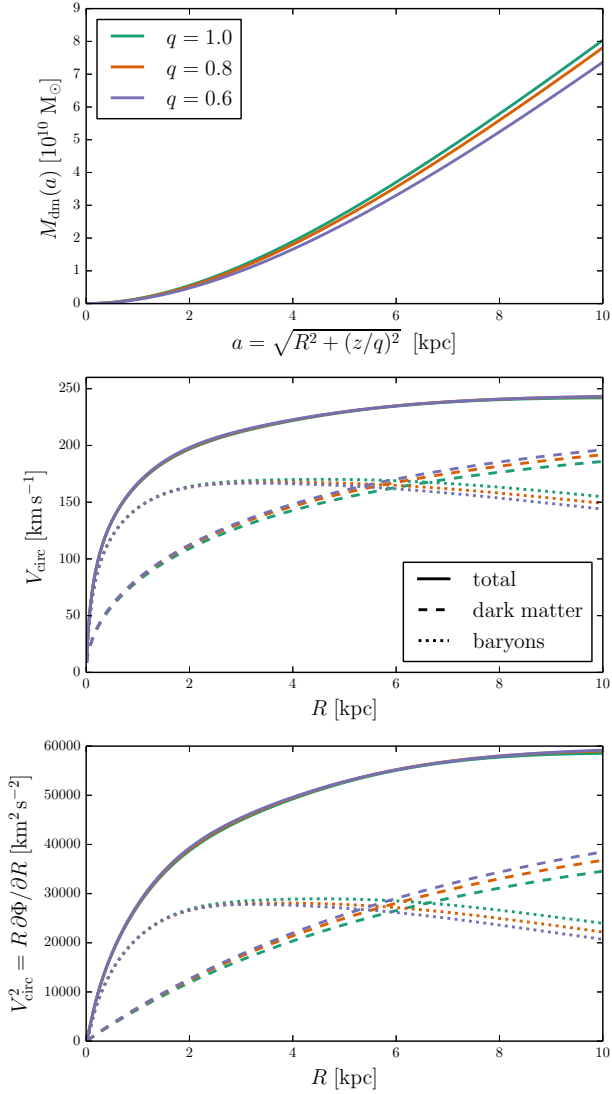}
 \caption{Upper panel: dark mass enclosed within the elliptical radius $a$ for
best-fitting models with three different axis-ratios $q$: 0.6, 0.8 and 1.
Middle panel: circular speed curves for the same models as in the upper
panel. The dashed and dot-dashed lines show the circular speed curve
generated by the dark matter halo and baryonic components alone. Bottom panel:
same as the middle panel but showing squared circular speed as a function of
$R$, which better illustrates the relative contributions of the dark
matter and the baryons to the radial force at each radius.}
 \label{fig:EllipsoidalMass}
\end{figure}
The inverse dependence of $\rhodm$ on $q$ implies that for similar scale
radii $r_{0,\rm dm}$ the mass of the dark matter halo within an oblate volume
with axis ratio $q$ is approximately independent of $q$. This is confirmed by
\figref{fig:EllipsoidalMass} (upper panel), which shows the cumulative mass
distribution as a function of elliptical radius.

The invariance of the dark matter mass profile can be qualitatively
understood by the following consideration: flattening the dark halo at fixed
local density reduces its mass and its contribution to the radial force,
$K_R$. But -- due to its still large thickness -- its contribution to the
vertical force $K_z$ at low $z$ remains almost constant or slightly grows. To
restore the value of the circular speed at the Sun's location we have to either increase
the mass of the halo or that of the disc. However, filling the gap with disc
material  increases $K_z$ and
consequently compresses the vertical mass profile predicted by the \df. Thus
the only possibility is to increase the mass of the halo and decrease the
mass of the disc in order to keep $K_z$ at the appropriate level. The upper
panel of \figref{fig:EllipsoidalMass} illustrates that this increase does not
fully balance the decrease from the initial flattening. The differences between the
lines is much smaller than the variations coming from the systematic
uncertainties. Taking the latter into account we find for the mass of dark
matter interior to an ellipsoidal surface that goes through the solar annulus
 $$
 M_{\rm dm}(a<R_0) = (6.0^{+1.35}_{-1.2}) \times 10^{10} \msun
$$
with $a = \sqrt{R^2 + (z/q)^2}$.

Interestingly, despite the fact that the enclosed dark mass is
slightly decreasing with decreasing axis ratio, its contribution to the
radial force increases. This is due to the enhanced gravitational pull of
a flattened distribution and leads to a smaller baryonic mass required to
sustain the rotation curve (middle and bottom panel of
\figref{fig:EllipsoidalMass}).

\begin{figure}
 \centering
 \includegraphics[width=0.47\textwidth]{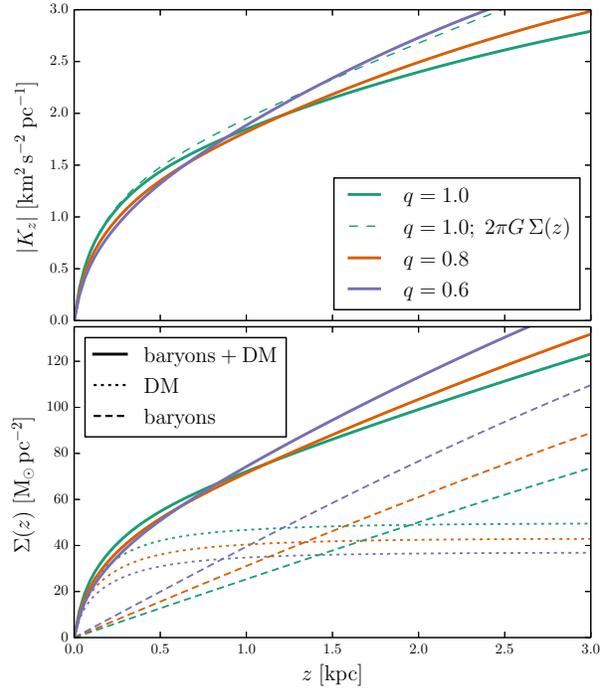}
 \caption{Upper panel: total vertical gravitational force as a function of
distance from the Galactic plane for best-fitting models with three different
dark halo axis-ratios $q$: 0.6, 0.8 and 1. The dashed turquoise line shows the
$K_z$ profile naively computed from the surface density profile using the
simple conversion often used in the literature. At $|z|>1\kpc$ this diverges
significantly from the true curve, namely the solid red line for the
spherical dark halo. Lower panel: same as above, but this time the surface
density $\Sigma(z)$ is shown: solid lines show the total surface density
while dashed and the dot-dashed lines show the contributions of the dark halo
and the Galactic disc.}
 \label{fig:flattening-Kz-SD}
\end{figure}

Fundamentally, the mass profiles are nearly independent of axis ratio because
the observed kinematics determine the ability of stars to resist $K_z$, and
the \citetJuric\ density profile shows the extent to which they do resist
$K_z$. Hence $K_z$ is narrowly constrained by the data independently of what
mass distribution generates it. $K_z$ is closely related to the surface
density
\[
 \Sigma(z) = \int_{-z}^z \d z' \rho(R_0,z').
\]
\figref{fig:flattening-Kz-SD} illustrates the $z$-dependencies of these
quantities for models with varying $q$. We see that there is a region around
$z=900\pc$ where all profiles intersect. Taking into account the systematic
uncertainties given above we find
$$
 |K_z(z=0.9\kpc)| = (1.78 \pm 0.4) \km^2\sec^{-2} \pc^{-1}.
$$
For the surface density between $\pm 900\pc$ we find
$$
  \Sigma(z=0.9\kpc) = (69 \pm 15) \msun \pc^{-2}.
$$
Below in \figref{fig:SD} we set these measurements in context with  estimates from the literature.

\subsection{Other properties}

We now give results for the model with a spherical dark halo. The
best-fitting model has a virial mass\footnote{We define the virial
mass as the mass interior to the radius $R_{200}$ that contains a mean
density of 200 times the critical density for a flat universe, $\rho_{\rm
crit}$.} $M_{200} = (1.3 \pm 0.1) \times 10^{12}\msun$. The above-mentioned
systematic uncertainties translate into a $<10$ per cent uncertainty in the virial
mass, but this does not encompass the uncertainty introduced by the assumed
shape of the radial mass profile of the dark-matter halo.
For the models with flattened halos we find slightly increased virial masses of $1.4 \times 10^{12}\msun$ and $1.6 \times 10^{12}\msun$ for the axis ratios 0.8 and 0.6, respectively.

The total mass of the Galaxy's stellar disc is $(3.7 \pm 1.1) \times10^{10}\msun$. This is lower but not far from the canonical value of $5\times 10^{10}\msun$. It is within the range of 3.6 -- 5.4 $\times10^{10}\msun$ estimated by \citet{Flynn2006}.
Combining the  stellar disc with the bulge and the gas disc, we arrive at a total baryonic
mass $(5.6 \pm 1.6) \times 10^{10}\msun$, or a baryon fraction $(4.3 \pm 0.6)$ per cent.
This value is much lower than the cosmic baryon fraction of $\sim 16$ per cent \citep{WMAP-9,Planck2013-XVI}, once again
illustrating the ``missing baryon problem'' \citep[e.g.,][]{Klypin1999}. While this baryon fraction
does not include the mass of the Galaxy's virial-temperature corona, the mass
of the corona within $\sim20\kpc$ of the Galactic centre is negligible
\citep{Marinacci2010}; the missing baryons have to lie well outside the visible
Galaxy in the circum- or inter-galactic medium.

The thick disc contributes about 32 per cent of the disc's stellar mass which is lower than the 70 per cent found by \citetJuric. This result
depends, however, on our decision to equate the radial scale lengths of the
two discs. If the scale length of the thick disc is assumed to be shorter,
as found by \citet{Bovy2012c}, the mass fraction in this component increases
to $\sim 60$ per cent. The better agreement with \citetJuric\ is only apparent, however, because these authors found a
\emph{longer} scale radius for the thick disc.

\begin{figure}
 \centering
 \includegraphics[width=0.47\textwidth]{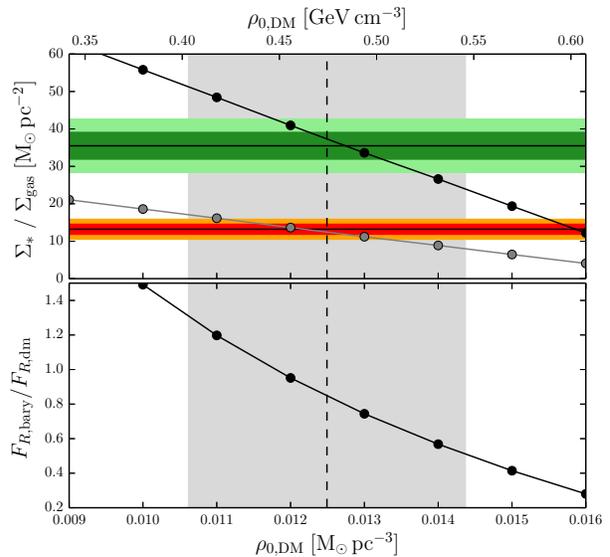}
 \caption{Upper panel: mass surface densities in our models for the stars
(black points and lines) and gas (grey points and lines). The green and
orange shaded area show the corresponding one/two sigma regions reported by
\citet{Flynn2006}. Lower panel: the ratio $F_{R,\rm bary}/F_{R,\rm dm}$ of
the contributions to the radial force at $R_0$ from baryons and dark matter.
In both panels the grey shaded area illustrates the systematic uncertainties
of $\rhodm$ with the (interpolated) best-fitting value marked by the black
dashed line. For this value we have $F_{R,\rm bary}/F_{R,\rm dm}\sim0.85$.}
 \label{fig:ForceRatios}
\end{figure}

\figref{fig:ForceRatios} shows for several fairly
successful spherical models the surface densities of the stellar and gaseous
discs at $R_0$ (upper panel) and the ratio of the radial forces at $R_0$ from
the baryons and dark matter (lower panel). The upper panel shows good
agreement with the
estimates of the baryonic surface densities derived from Hipparcos data by
\citet{Flynn2006} (coloured bands). The lower panel shows that equal
contributions to the radial force are achieved for local dark-matter
densities $\rhodm$ that are lower than our favoured value for a spherical halo, but
still within the range encompassed by the systematic uncertainties, which is
shaded grey.
In our best-fit model the solar neighbourhood is mildly dark-matter dominated
with only 46 per cent of the radial force coming from gas and stars.
Alternatively, we can look at the contribution of
disc to the total rotation curve at 2.2 times the scale radius to check
whether our disc is ``maximal'' according to the definition of
\citet{Sackett1997}. We find a ratio $V_{\rm c,disc}/V_{\rm c,all} = 0.63$
($V_{\rm c,baryons}/V_{\rm c,all} = 0.72$) that is below the range of 0.75 --
0.95 for a maximal disc, but slightly above the typical range of
$0.47\pm0.08$ ($0.57\pm0.07$) for external spiral galaxies
\citep{Bershady2011,Martinsson2013}. It is still lower than the value of
$0.83\pm0.04$ found by \citet{BovyRix2013}.

\section{Kinematics} \label{sec:kinematics}

Here we discuss the kinematic properties of our best-fitting model. The
circular speed at the solar radius, $v_{\rm c}(R_0)=240\kms$ is largely the
result of the adopted values of $R_0=8.3\kpc$, the proper motion of Sgr\,A*,
and $\vV_\odot$, the solar motion w.r.t.\ to the LSR. Our constraints for the
mass model actually fix the ratio $v_{\rm c}(R_0)/R_0$ \citep{McMillan2011}.

For the local escape speed $v_{\rm esc} = \sqrt{2\Phi(R_0)}$ we find a value
of $613\kms$. \citet{Piffl2014} recently found a lower value of
$533^{+54}_{-41}\kms$, but for this they used a modified definition of the
escape speed as the minimum speed needed to reach $3R_{\rm vir}$. If we apply
their definition to our model we find a value of $580\kms$ which is still on the high
side, but within their 90 per cent confidence interval. The uncertainties arising
from the above mentioned systematics on this value are of order 1 per cent. This
comes mainly from our rather strong prior on the mass within 50~kpc and again
does not cover the uncertainties in the dark matter profile at large radii\footnote{Because of this and also because of the focus of \citet{Piffl2014} on the fastest stars in the RAVE survey, which carry most of the information on the escape speed, we still consider their value as the more robust one.}.

\begin{figure*}
 \centering
 \includegraphics[width=0.98\textwidth]{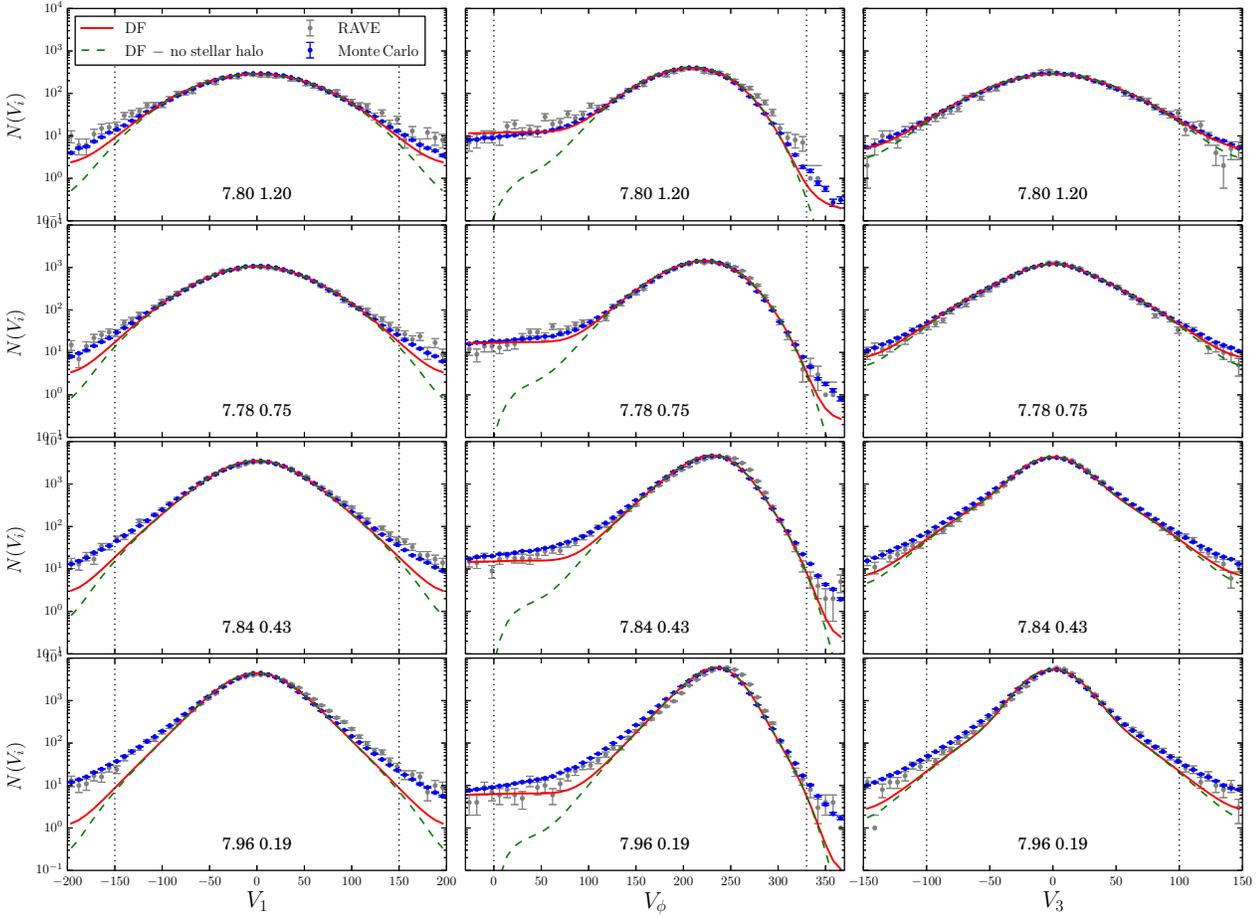}
 \caption{Velocity distributions in four spatial bins. In each panel the
numbers at the lower centre give the $(R,z)$ coordinates of the bin's
barycentre. The grey error bars show the RAVE data, while the green error
bars show the mean of ten Monte Carlo re-samples of our model with $\rhodm =
0.012\msun\pc^{-3}$, which is close to our best-fit value. The vertical
dotted lines show the boundaries of the velocity ranges over which $\chi^2$
was computed. The full red curves show the predictions of the \df\ in the absence
of distance errors, while the dashed lines show the corresponding predictions
when the \df\ of the stellar halo is deleted. }
 \label{fig:giants}
\end{figure*}

The data points in \figref{fig:giants} show histograms for each principal
velocity component and spatial bins defined by $7.3\kpc <R <R_0$ and ranges
in $z$ that increase from bottom to top: the upper limits of the bins are
at $z=0.3,0.6,1,1.5\kpc$ and the coordinates of each bin's barycentre are given
at the lower centre of each panel. The vertical scales of the plots are
logarithmic and cover nearly three orders of magnitude in star density. The
plotted velocity components $V_1$ and $V_3$ are along directions $\ve_1$ and
$\ve_2$ that lie very close to the radial and latitudinal directions,
respectively, their precise directions being those determined by
\citet{Binney2014b} to be the eigenvectors of the velocity dispersion tensor.
The Poisson error on each data point is marked by the error bar and is in
most cases insignificant. The grey data points represent the original RAVE
data, while the green symbols show the mean of 100 Monte Carlo re-samplings
using the mass model--\df\ pair with $\rhodm=0.012\msun\pc^{-3}$, which is
closest to our best-fit value. Within the considered velocity ranges
(bracketed by the vertical dashed lines in the panels) the
fits to the data are good. For the $V_\phi$-distributions there is a systematic shift
by approximately $5\kms$. This shift is much smaller or even reversed for the
spatial bins that lie beyond $R_0$, which suggests that it arises from
spiral structure.

Far from the plane, the wings of the $V_1$ distributions are systematically
under-predicted while there is a hint of an over-prediction near the plane. A
similar trend is apparent for the low-angular momentum stars in the $V_\phi$
distributions and this anomaly is also present in the histograms for bins
that lie outside $R_0$. These trends suggest that the stellar halo
density decreases less quickly with increasing $|z|$ than our model predicts, and hence that this component
is \emph{less} flattened than we have implicitly assumed by adopting the form
(\ref{eq:haloDF2}) of the function $h$ that appears in the halo \df.
A definitive statement is, however, impossible,
because the wings of the distribution are strongly affected by the uncertain
distance estimates (and our modelling of this effect). Stars at greater
heights from the Galactic plane, such as those in the SEGUE survey
\citep{SEGUEpaper} would yield a clearer picture.

\begin{figure*}
 \centering
 \includegraphics[width=0.98\textwidth]{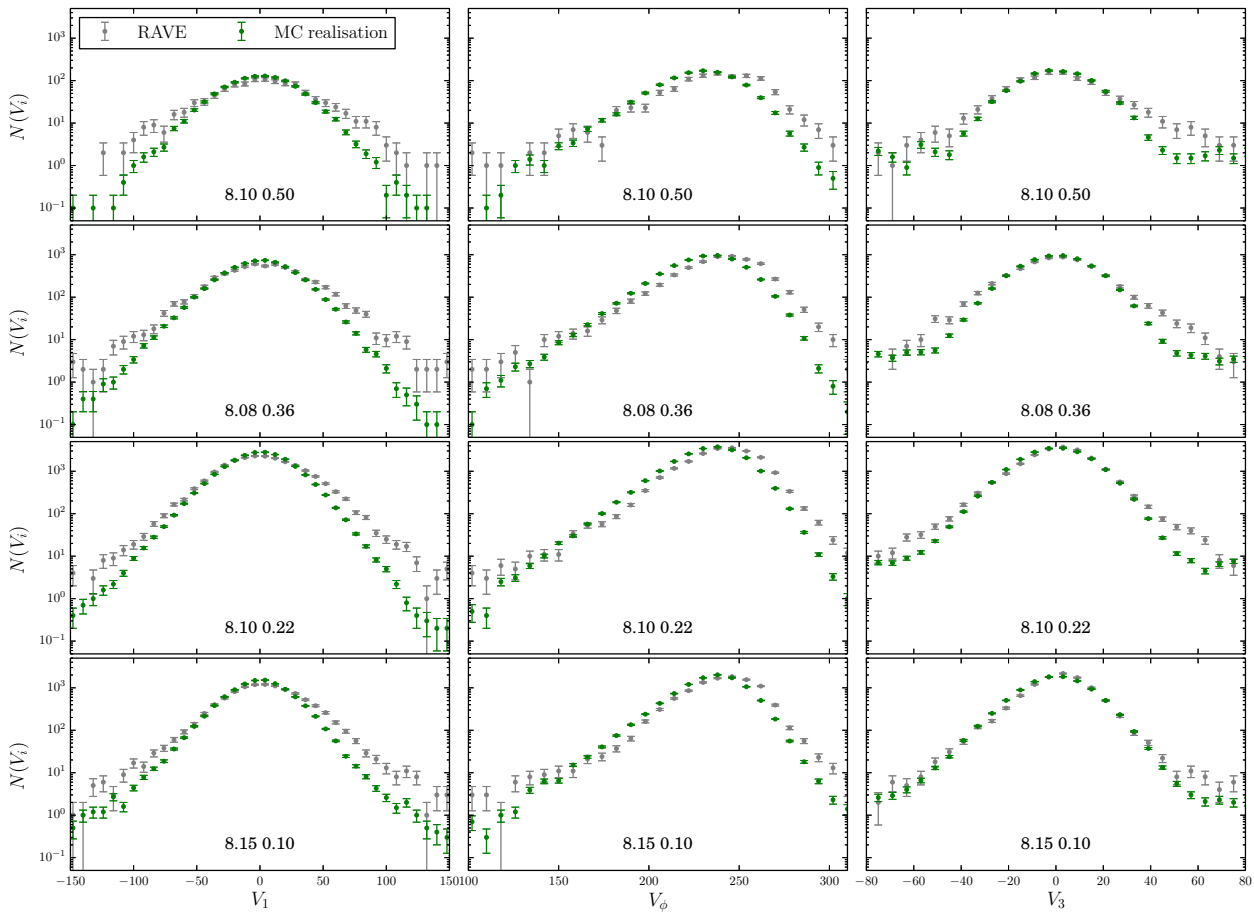}
 \caption{Velocity distributions of hot dwarfs in four spatial bins. These
plots are similar to those in \figref{fig:giants}, but the  borders of the
bins are (from bottom to top) at $|z|=0.15,0.30,0.45$ and $0.6\kpc$. The
$(R,z)$ positions of the barycentres of the sub-samples are given in each
panel. The grey error bars show the RAVE data while the green error bars show
the mean of ten Monte Carlo re-samples of the model with $\rhodm =
0.012\msun\pc^{-3}$, which is close to our best-fit value.}
 \label{fig:hot_dwarfs}
\end{figure*}

\subsection{Kinematics of hot dwarfs}
We next look at the kinematics predicted for a different set of RAVE
observations, namely hot dwarf stars. Stars in this group, defined by $\logg
> 3.5\dex$ and $\Teff > 6000\K$, have the most precise distance estimates,
but due to their lower luminosity they can be seen in a smaller
volume than the giants.  So we use spatial bins defined by smaller distances
from the Galactic plane: we place the borders at $|z|= 0.15, 0.3, 0.45,
0.6\kpc$.  Since these stars cannot be old, we assume their \df\ is given by
the portion of the thin-disc \df\ for age below $5\Gyr$ ($=\tau_{\rm m}$).
The green data points in \figref{fig:hot_dwarfs} show the resulting Monte
Carlo re-sampling of stellar distances and velocities, while the grey points
show the RAVE data. The predicted and observed distributions agree well in
their cores, but in their wings the RAVE data lie higher, especially in the
case of $V_1$. The predicted $V_\phi$ distributions under-populate the
high-$V_\phi$ tails. Both phenomena are symptomatic of a model population
that has less random motion than the real one. It might be argued that
$5\Gyr$ is an excessively young age to adopt for this population, which will
contain stars with masses down to $1.1\msun$ that live in excess of $7\Gyr$.
With an age cutoff of $7.5\Gyr$ improved fits to the $V_1$ distributions are
obtained, but overall the velocity distributions remain deficient, and anyway the
adopted age cutoff should match the mean lifetime of the population, not that
of the longest-lived stars within it.

The discrepancies between theory and observation in \figref{fig:hot_dwarfs}
tie in with the \df's curve for the thin disc in \figref{fig:all_zprofs2} falling rather steeply at
large $|z|$. It also ties in with a surprising result encountered by
B12b when fitting \df s like those used here to the GCS: the
thin-disc \df\ fully populated the wings of the $U$ distribution of GCS stars
with the result that when a thick-disc \df\ was added, using both the GCS
velocities and the vertical density profile of \citet{Gilmore1983}, the
implied radial velocity dispersion of the thick disc was lower than that of
the old thin disc. Here bins at large $|z|$ strongly influence the fitted
\df, so the \df\ of the thick disc has sufficient in-plane dispersion to fill
out the wings of these distributions. With our chosen functional form of the
\df, the thick disc's \df\ then places  enough stars in the bins near the plane to fill
out the wings of these distributions. It follows that the thin-disc \df\ is
then chosen such that it does not place additional, now unwanted, stars in the
wings of the $V_1$ distributions at low $|z|$. Unfortunately, the RAVE
distribution for hot dwarfs, which are surely thin-disc stars, shows that the
thin disc {\it does} make a non-negligible contribution to the wings of the
$V_1$ distribution.

The difficulty encountered here with the hot dwarfs and the difficulty
encountered by B12b with the thick disc may both arise from
an inappropriate choice for the functional dependencies of the \df\ on $J_r$ and
$J_z$. These dependencies are not motivated by any convincing physical
arguments \citep[but see][for a relevant discussion]{BinneyTeneriffe}, they
are simply those provided by familiar analytic functions that have the right
general properties. The superb statistics provided by the RAVE survey may
oblige us to tweak these functions. This is, however, a topic for another
paper.

%%%%%%%%%%%%%%%%%%%%%%%%%%%%%%%%%%%%%%%%%%%%%%%%%%%%%%%%%%%%%%%%%%%%%%%%%%%%%%%%%%%%%%%%%%%%%%%%%%%%%%%%%%%
% 
\section{Discussion} \label{sec:Discussion}
We have built models of the Galaxy using the vast data set provided by the
RAVE survey together with additional observations constraining the Galactic
rotation curve and the stellar density profile above and below the Sun. In
combination these measurements allowed us to disentangle the contributions to
the local gravitational field of the baryonic and the dark matter and hence
put tight constraints on the mass of the dark halo that lies interior to the
Sun, and on the surface density of matter that lies within $\sim0.9\kpc$ of
the plane near the Sun. In order
to compare our results to previous measurements in the literature one should
keep in mind that our quoted uncertainties are not the statistical
uncertainties which are negligible ($<0.4$ per cent), but originate purely from a
conservative estimate of our systematic uncertainties. Similar systematic
uncertainties are generally present in other studies.

A comprehensive review of estimates of the local dark matter density was
recently published by \citet{Read2014}. Most of the more recent studies find
lower values of $\rhodm$ than our estimate of
$0.0126 \msun \pc^{-3} \pm 15$ per cent for a spherical dark halo. Almost
all previous studies rely on the assumption that the motions of stars near
the Sun are separable into radial and vertical components.
Recent examples of this approach are
\citet{BovyTremaine2012} ($0.008 \pm 0.003\msun \pc^{-3}$) or
\citet{Zhang2013} ($0.0065 \pm 0.0023 \msun \pc^{-3}$). However,
\citet{Garbari2012} showed by analysing mock data sets from $N$-body
simulations that this can bias the results to lower values. They propose a
new method that weakens this assumption and use this to re-analyse the data
by \citet{Kuijken1991}. Both their estimate and
their revised value $0.0087^{+0.007}_{-0.002}\msun\pc^{-3}$ in
\citet{Read2014}, are consistent with our results.
Also most other recent studies that do not rely on
the separability assumption find value for $\rhodm$ that are in good
agreement with our measurement \citep{McMillan2011, Salucci2010, Iocco2011,
Nesti2013}.

Parallel to this work \citet{Bienayme2014} used a sample of $\sim 4600$ red
clump stars from RAVE to measure the vertical force (see \figref{fig:SD}
below) and from a more classical approach concluded that the local
dark-matter density is $\rhodm = 0.0143\pm0.0011\msun\pc^{-3}$. Their results
are very similar to the results of this study despite a radically different
methodology, different distance estimates, and only a partial overlap of
the stellar samples. In particular, they used stars at $|z|>1.5\kpc$ that were
excluded for our study.
\subsection{Surface density and vertical force}
\begin{figure}
 \centering
 \includegraphics[width=0.47\textwidth]{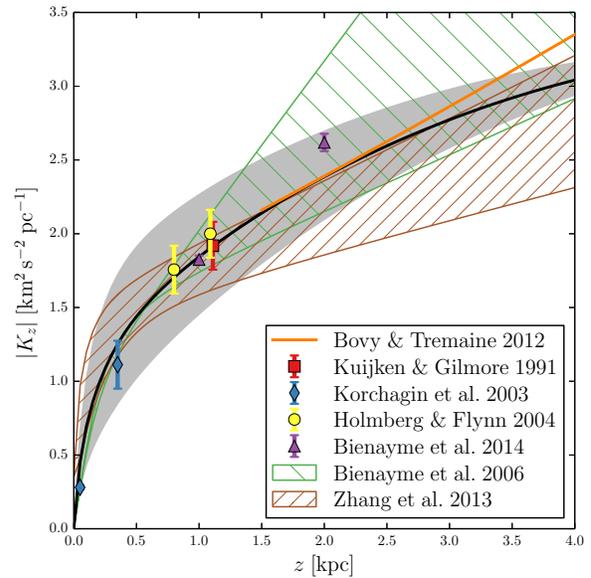}
 \caption{Vertical force law for our best-fitting model with a spherical dark
halo (solid black line) and its systematic uncertainty (grey shaded area).
The error bars and hatched regions show several other estimates from the
literature. The estimate by \citet{BovyTremaine2012} was computed using their preferred parameters (dashed curve in their Figure~2).}
 \label{fig:SD}
\end{figure}
It is often assumed that once one has determined the vertical force $K_z$
acting on stars at a certain height $z$ above or below the Galactic plane,
this can be trivially converted into the surface mass density between $\pm
z$. It was already remarked before that this is a good approximation only for
small $|z|$. A comparison of the dashed and solid red lines in the upper panel
of \figref{fig:flattening-Kz-SD} illustrates this once again \citep[see also][]{BovyTremaine2012}. 
Previously, with sample sizes of hundreds or few thousand stars, the distinction could be
generously ignored. With almost 200\,000 stars the statistical uncertainties
become negligible and adopting $K_z=2\pi G\Sigma$ introduces a significant
bias. It is hence advisable not to confuse $K_z$ and $2\pi G\Sigma$.

The full black curve in \figref{fig:SD} shows the $K_z$ profile of our
best-fitting model, while the grey band shows our systematic uncertainty.
Clearly these results are in excellent agreement with the data points, which
show the  estimates of \citet{BovyTremaine2012},
\citet{Kuijken1991}, \citet{Korchagin2003}, \citet{Holmberg2004},
\citet{Bienayme2006, Bienayme2014} and \citet{Zhang2013}. Note, that
uncertainties of these earlier studies are random errors, while our uncertainties
from random errors are negligible.

Interestingly, the study by \citet{Bienayme2014} extended the range of $|z|$
values probed to $2\kpc$, so above the volume in which our stars reside.
Their steep gradient between 1 and $2\kpc$ could be interpreted as
indicating a flattened dark halo with axis ratio $q \lesssim 0.8$ (cf.\ the upper
panel of \figref{fig:flattening-Kz-SD}). Their slightly higher value for
$\rhodm$ is consistent with this, as it requires $q$ to be 0.79--0.94 to
match our results.

\citet{MoniBidin2012b} claimed that the kinematics of a sample of distant
red giants from \citet{MoniBidin2012a} proves that $\rhodm$ is negligible.
However, \citet{Sanders2012} showed that the velocity dispersion gradients
reported by \citet{MoniBidin2012a} were most likely too shallow by a factor
2--3 and their uncertainties severely under-estimated.
\citet{BovyTremaine2012} showed that the analysis in \citet{MoniBidin2012b}
is flawed, and from the data in \citet{MoniBidin2012a} derived a value for
$\rhodm$ that is consistent with a significant dark halo. However, the
errors on $\rhodm$ given by \citeauthor{BovyTremaine2012} are based on the
erroneous errors in \citet{MoniBidin2012a}, and the use of realistic errors
would probably prohibit any significant statement.

We nevertheless include the results of
\citet{BovyTremaine2012} in \figref{fig:SD}. Surprisingly,
\citeauthor{BovyTremaine2012} find a steeper force gradient than our standard
model, yet derive a smaller dark matter density from it. They underestimate
the surface density (cf.\ \figref{fig:flattening-Kz-SD}), and hence the local
dark matter density, because they use the over-simple conversion from $K_z$
to density mentioned above.

\begin{figure}
\centering
 \includegraphics[width=0.47\textwidth]{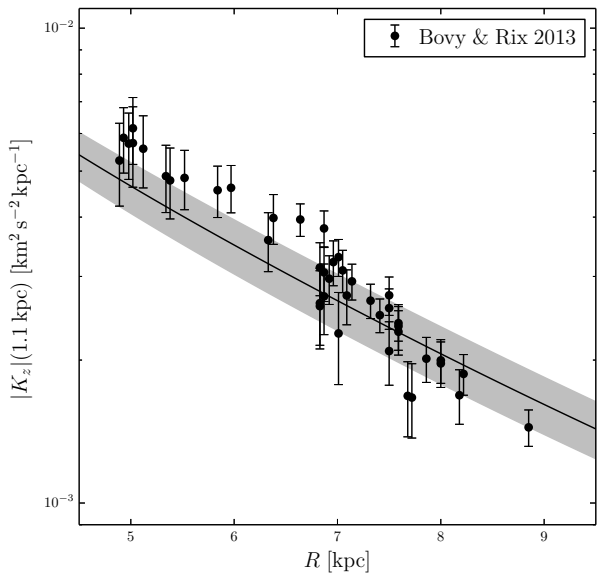}
 \caption{Data points: estimates of $K_z(R, 1.1\kpc)$ obtained by \citet{BovyRix2013}
 for mono-abundance populations within the SEGUE survey. The black line and the grey-shaded area indicate our best-fit model and uncertainty.}
 \label{fig:BovyRix}
\end{figure}

\citet{BovyRix2013} used quasi-isothermal \df s to derive the estimates of $K_z$
plotted in \figref{fig:BovyRix} from several populations within the sample of
G dwarfs studied by \citet{Lee2011b} as part of SEGUE. Each ``mono-abundance''
population comprises stars that lie in a small bin in the
$(\alpha/\hbox{Fe}/[\hbox{Fe/H}])$ plane and consists of a couple of hundred
stars at most. \citet{BovyRix2013} derived an independent gravitational potential
and quasi-isothermal \df\ for each sub-population by requiring that the
population's vertical kinematics and spatial distribution are consistent with
the observations.  They argued that each population constrained $K_z$ at a
particular radius, and each point in \figref{fig:BovyRix} shows the
constraint provided by one population at the corresponding radius. Since the
chemical information we have for our much larger sample of stars is
associated with large uncertainties, and we are sceptical that individual
mono-abundance populations have quasi-isothermal \df s, we have used a much
simpler, metalliticity-blind, approach to the data in which it is fundamental
that all stars move in the same gravitational potential. Nonetheless, in the
region $R>6.6\kpc$, where the available data are most clearly relevant, the
Bovy \& Rix points fall nicely within our region of systematic uncertainty,
shaded grey in \figref{fig:BovyRix}.
\subsection{Structural parameters of the discs}
Our values for the parameters defining the vertical mass profile of the
Galactic disc are systematically lower than the canonical values of $300\pc$
for the thin disc and 900 to  $1400\pc$ for the thick disc
\citep[J08]{Gilmore1983}. One should keep in mind, however, that these values, together
with the normalisation parameter $\fthick$, are not fully constrained by the
data and their uncertainties. It is well known, that extremely similar
profiles are produced by parameter values that fall in an extended region of
parameter space (e.g.\ \citetJuric).

With our model we do not suffer from this degeneracy in the sense that our
values for $\zthin$ and $\zthick$ are fixed during the alignment of the
\df\ and the mass model, which is done by comparing smooth curves that have no
uncertainties. Since we are obtaining good fits to the \citetJuric\ data (and
the \citet{Gilmore1983} data), there can necessarily be no significance in
any difference of the parameters. 

\subsection{Baryonic mass content}
\citet{Flynn2006} find the total luminosity surface density at the Sun in the
$I$-band to be $29.5\Lsun\pc^{-2}$ with an uncertainty of 10 per cent.
Combining this with our results we find a total luminosity of the stellar
disk of $(3.95 \pm 0.15) \times 10^{10} \Lsun$. With an additional
contribution of $10^{10} \Lsun$ from the bulge \citep{Flynn2006, Kent1991} we
arrive at a total luminosity $L_{I} \simeq 5\times10^{10}\Lsun$ ($M_I \simeq
-22.4$) and an $I$-band mass-to-light ratio of $(1.4 \pm 0.2)$.
These values are very similar to analogous
estimates by \citet{Flynn2006} ($M_I \sim -22.3$, $(M/L)_I \sim 1.3$) and
\citet{BovyRix2013} ($M_I \sim -22.5$, $(M/L)_I \sim 1.3$).
\subsection{Consistency with $\Lambda$CDM}
During our model fitting process we apply a prior in the concentration
parameter $c$ of the dark matter halo that was based on cosmological simulations 
of structure formation (Section~\ref{sec:concentration_prior}).
There are, however, several other constraints that have to be satisfied and our
best fitting model has a concentration $\ln(c)=3.0$ that is almost $3\sigma$
above
the central value of the prior. A similar result was already obtained by
\citet{McMillan2011}. This illustrates that the additional information
from this prior was over-ruled and, indeed, our results do not change
significantly when we use a flat prior instead.

At first glance this seems to point to a mild tension with the
predictions from $\Lambda$CDM cosmology. One has to keep in mind, however,
that the above mentioned simulations were dark-matter only runs. The
formation of galaxies in the centres of the dark matter haloes will alter
their shapes and radial profiles. The classic halo response model of
\citet{Blumenthal1986}, which assumes an adiabatic contraction of the radial
mass profile, is now thought to over-predict the response
\citep[e.g.][]{Gnedin2004, Gnedin2010, Abadi2010}.  There is also general
agreement that there is no universal response that depends only on the final
distribution of the baryons, but that the response also depends on the
specific accretion history of the galaxy \citep{Abadi2010, Gnedin2010}.

\begin{figure}
 \centering
 \includegraphics[width=0.47\textwidth]{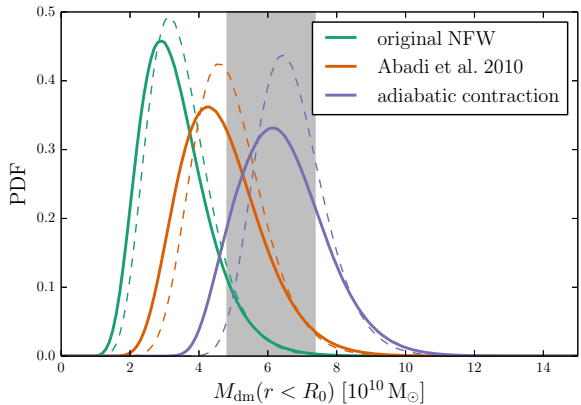}
 \caption{PDFs for $M_{\rm dm}(r<R_0)$ in the $\Lambda$CDM cosmology for
 galaxies with virial masses between 0.5--2 $\times 10^{12}\msun$. The different
 lines show the predictions when different degrees of contraction due to the
 formation of central galaxies in the dark haloes are taken into account. The
 grey shaded area marks the result of our study. The dashed lines show the
 predictions for a virial mass equal to our best-fit value. These are shifted to higher value, because our best-fit value of $1.3\times10^{12}\msun$ is slightly on the high side of the above mass range.}
 \label{fig:LCDM-M_in}
\end{figure}

Instead of looking at the global shape of the halo, we can compare the
predictions of numerical simulations with our robust result on the dark mass
contained within the solar radius, $M_{\rm dm}(r<R_0)$ -- a similar approach
was taken by \citet{Navarro2000a} and \citet{Abadi2010}. We
use the mass-concentration relation reported by \citet{Maccio2008},
\begin{equation} \label{eq:Mc-relation}
 \log c_{200} = 0.917 - 0.104\log(M_{200}~[10^{12}h^{-1}\msun]),
\end{equation}
with $h=0.73$, to compute $M_{\rm dm}(r<R_0)$ as a function of halo mass
$M_{200}$ as follows.  Given $M_{200}$, we find
$c_{200}$ from equation~(\ref{eq:Mc-relation}) and compute the mass interior
to $R_0$ in a standard NFW model. For the contracted profiles, we first
tabulate the standard NFW mass profile and then contract it according to the
prescriptions of \citet{SW1999} (adiabatic) and \citet{Abadi2010} using the
baryonic disc and bulge configurations from our best-fit model. The mass
interior to $R_0$ can then be obtained by interpolation. The cosmic
scatter around the relation~(\ref{eq:Mc-relation}) is well approximated by a
log-normal distribution and \citet{Maccio2008} find $\sigma_{\log c} = 0.105$
and hence we obtain a probability distribution for $M_{\rm dm}(r<R_0)$
for each $M_{200}$. We can then integrate over a
plausible range of Milky Way masses to obtain a probability distribution for
$M_{\rm dm}(r<R_0)$ for our Galaxy. We choose a flat prior between 0.5 and
2~$\times 10^{12}\msun$ that roughly covers the range of Milky Way masses
reported in the literature. \figref{fig:LCDM-M_in} shows the results.  If we
assume the original NFW profile we find our model to be a mild outlier as
expected from the stronger concentration.  Note that in contrast to
\citet{Abadi2010} we find that the Milky Way has \emph{more} mass inside
$R_0$ than predicted by cosmology. If we modify the mass profile via
adiabatic contraction or the prescription for an intermediate contraction
advocated by \citet{Abadi2010}, we find much better agreement. The old age of
the Galaxy's thin disc implies that our Galaxy has avoided significant
mergers for $>8\Gyr$. \citet{Abadi2010} speculated that in such a case the
dark halo would contract more strongly than predicted by their formula. We
also note, that \citet{Gnedin2004, Gnedin2010} generally predict a
contraction that is closer to the adiabatic case than \citet{Abadi2010}.
Hence, when halo contraction is taken into account we are in excellent
agreement with the Milky Way being a typical spiral galaxy in a $\Lambda$CDM
universe.
%
%%%%%%%%%%%%%%%%%%%%%%%%%%%%%%%%%%%%%%%%%%%%%%%%%%%%%%%%%%%%%%%%%%%%%%%%%%%%%%%%%%%%%%%%%%%%%%%%%%%%%%%%%%%
%
\section{Conclusions}
We have explored the vertical profile of mass density within $\sim1.5\kpc$ of
the Galactic plane by combining the kinematics of $\sim200\,000$ giant stars
in the RAVE survey with
estimates of the variation with $z$ in the number density of disc stars
determined by \citet{Juric2008} (\citetJuric) from the SDSS. We have done this using a novel
approach to dynamics in which for a given mass model we fit a parametric \df\
of the form $f(\vJ)$ to the kinematics, and then compute the
vertical density profile predicted by the chosen \df. This process is repeated
using different mass models until the predicted stellar density profile
agrees with the observed one. 

That we are able to obtain excellent agreement between the predicted and
observed stellar profiles is a non-trivial result, and suggests (a) that the
parametric form of the \df\ with which we have chosen to work is capable of
approximating the true \df\ quite accurately, and (b) that our mass models
have an appropriate form. The latter comprise double-exponential discs for
the ISM, the thin and thick stellar discs plus a NFW dark-matter halo and a
bulge with spheroidal equidensity surfaces.

While the data we use here constrain tightly the vertical structure of the
disc, other data constrain the Galaxy's radial structure better.
Consequently, we have determined the radial scale lengths of the density
profiles of both the discs and the dark halo from measurements of the proper
motion of Sgr A*, terminal velocities of gas inside the Sun, and
six-dimensional phase-space coordinates of maser sources. Our models fit
these data to well within the errors. The radial scales $R_\sigma$ on which
the velocity-dispersion parameters $\sigma_r$ of and $\sigma_z$ of the thin
and thick discs decrease with radius have to be extracted from the RAVE data.
For the thick disc the RAVE data yield a reasonable value of $R_\sigma$, but
they do not yield credible values of $R_\sigma$ for the thin disc because the
survey does not probe the thin disc over a large enough radial range.
The low-latitude infra-red survey APOGEE
\citep{APOGEEpaper} should  resolve this issue, but to date it has not
yielded physically plausible results \citep{Bovy2012b}.
Consequently, we have simply imposed $R_\sigma=9\kpc$ to ensure that the thin
disc has a roughly radius-independent scale height.

The star samples we are using are so large that statistical errors are
insignificant, so the uncertainties in our results are entirely systematic. By far
the biggest source of uncertainty are the solar distance to the Galactic centre and systematic errors in the distances
employed in the determination of both the kinematics and the density profile.
\citetJuric\ used a colour-magnitude relation to determine the density of M
dwarfs stars. The profile constructed in this way, which is in excellent agreement with the original \citet{Gilmore1983}
profile, is expected to be steeper than the true profile because
the observed stars are on average more luminous than they appear to be on
account of both Malmquist bias and undiagnosed binarity. We have allowed for
these effects by imposing them on our models, using the photometric errors
quoted by \citetJuric\ and a 10 per cent binary fraction. If one uses  distances
that are 20 per cent too small, one derives a local dark-matter density $\rhodm$ that is
$\sim8$ per cent too low. Similarly, decreasing the assumed value of $R_0$ from
$8.3$ to $8\kpc$ causes $\rhodm$ to decrease by $\sim5$ per cent.

What we are really measuring is the vertical profile of gravitating matter,
regardless of whether it is baryonic or dark. If we flatten an initially
spherical dark halo to axis ratio $q<1$ while holding constant its mass, the
contribution of the halo to the radial force on the Sun rises, as does the
halo's contribution to the vertical force that keeps disc stars near the
plane. Consequently, to keep our model consistent with the data, the mass of
the baryonic disc has to decrease if one decides to use a flattened halo. Hence
flatter halos require lower-mass baryonic discs. In fact, one finds that as the halo
is flattened, its mass also has to decrease slightly if the data are to be fitted
perfectly, so $\rhodm\propto q^{-0.89}$ rather than the relation
$\rhodm\propto q^{-1}$ that would apply if the dark-halo mass
were independent of $q$. The surface density within $|z|=0.9\kpc$ is
$69\pm15\msun\pc^{-2}$ independent of halo flattening.

With a spherical halo the total local surface density of the baryonic disc is
$37\msun\pc^{-2}$ in excellent agreement with the estimate of
\cite{Flynn2006}, and the baryons contribute 46 per cent of the radial force on the
Sun. The virial mass of the halo is $M_{200}=(1.3\pm0.1)\times10^{12}\msun$
and the mass of the Galaxy's baryons  is $(5.6\pm1.6)\times10^{10}\msun$, giving a
baryon fraction (stars and cold gas) as small as $(4.3\pm0.6)$ per cent. 

We have used our models to predict the kinematics of hot dwarf stars under
the assumption that these stars faithfully sample the contribution to the
thin-disc \df\ from stars with ages $<5\Gyr$. In the cores of the velocity
histograms, the predictions agree well with the RAVE data, but there is a
tendency for the wings to be under-predicted by the models, so the division
of the overall \df\ into thin- and thick-disc components may be imperfect, with
the young thin disc being slightly colder than it should be. On the other
hand, the models predict slightly lower mean-streaming velocities than implied by the
data, which is the opposite of what one would expect if the thin disc were
too cold. So in subsequent work it may be necessary to adopt a more complex
functional form for the \df.

The fundamental assumption underlying our work is that  RAVE stars sample the
same population as that probed by \citetJuric\ and \citet{Gilmore1983}. The
worrying aspect of this assumption is that we have used  giants from
RAVE, whereas \citetJuric\ and Gilmore \& Reid used dwarfs. We do have the
reassurance that \citet{Binney2014b} found the kinematics of cool dwarfs and
giants  to be consistent with one another in RAVE. However, it would be safer
to compare the kinematics of the RAVE giants to estimates of the density
profile that one derives from them, and we plan to do that soon. In this
connection it will be valuable to derive the density profile from a mixture
of the 2MASS and SDSS surveys since SDSS stars are too faint to sample the
thin disc well.

In this paper we have fitted models to the data in phase space rather
than the space of the observables (proper motions, parallaxes, magnitudes,
etc). More robust handling of errors is possible if one fits in the space of
observables \citep{McMillan2013}, and we hope to use the RAVE data in this
way soon.

\section*{Acknowledgements}
The research leading to these results has received funding from the European Research
Council under the European Union's Seventh Framework Programme (FP7/2007-2013)/ERC
grant agreement no.\ 321067.

Funding for RAVE has been provided by: the Australian Astronomical Observatory;
the Leibniz-Institut f\"ur Astrophysik Potsdam (AIP); the Australian
National University; the Australian Research Council; the French National Research
Agency; the German Research Foundation (SPP 1177 and SFB 881); the
European Research Council (ERC-StG 240271 Galactica); the Istituto Nazionale
di Astrofisica at Padova; The Johns Hopkins University; the National Science
Foundation of the USA (AST-0908326); the W. M. Keck foundation; the Macquarie
University; the Netherlands Research School for Astronomy; the Natural
Sciences and Engineering Research Council of Canada; the Slovenian Research
Agency; the Swiss National Science Foundation; the Science \& Technology
Facilities Council of the UK; Opticon; Strasbourg Observatory; and the
Universities of Groningen, Heidelberg and Sydney. The RAVE web site is at
http://www.rave-survey.org.

Figures~\ref{fig:flattening-Kz-SD}, \ref{fig:EllipsoidalMass}, \ref{fig:SD} and \ref{fig:LCDM-M_in} were produced using colour schemes from http://www.colorbrewer2.org.
\bibliographystyle{mn2e}
\bibliography{all_references}

\begin{thebibliography}{92}
\providecommand{\natexlab}[1]{#1}

\bibitem[{{Abadi} et~al.(2010){Abadi}, {Navarro}, {Fardal}, {Babul} \&
  {Steinmetz}}]{Abadi2010}
{Abadi} M.~G., {Navarro} J.~F., {Fardal} M., {Babul} A., {Steinmetz} M., 2010,
  \mnras, 407, 435

\bibitem[{{Allende Prieto} et~al.(2008)}]{APOGEEpaper}
{Allende Prieto} C. et~al., 2008, Astronomische Nachrichten, 329, 1018

\bibitem[{{Antoja} et~al.(2012)}]{Antoja2012}
{Antoja} T. et~al., 2012, \mnras, 426, L1

\bibitem[{{Aumer} \& {Binney}(2009)}]{Aumer2009}
{Aumer} M., {Binney} J.~J., 2009, \mnras, 397, 1286

\bibitem[{{Bershady} et~al.(2011){Bershady}, {Martinsson}, {Verheijen},
  {Westfall}, {Andersen} \& {Swaters}}]{Bershady2011}
{Bershady} M.~A., {Martinsson} T.~P.~K., {Verheijen} M.~A.~W., {Westfall}
  K.~B., {Andersen} D.~R., {Swaters} R.~A., 2011, \apjl, 739, L47

\bibitem[{{Bienaym{\'e}} et~al.(2006){Bienaym{\'e}}, {Soubiran}, {Mishenina},
  {Kovtyukh} \& {Siebert}}]{Bienayme2006}
{Bienaym{\'e}} O., {Soubiran} C., {Mishenina} T.~V., {Kovtyukh} V.~V.,
  {Siebert} A., 2006, \aap, 446, 933

\bibitem[{{Bienaym{\'e}} et~al.(2014)}]{Bienayme2014}
{Bienaym{\'e}} O. et~al., 2014, ArXiv e-prints, 1406.6896

\bibitem[{{Binney}(2010{\natexlab{a}})}]{Binney2010}
{Binney} J., 2010{\natexlab{a}}, \mnras, 401, 2318

\bibitem[{{Binney}(2012{\natexlab{a}})}]{Binney2012a}
{Binney} J., 2012{\natexlab{a}}, \mnras, 426, 1324

\bibitem[{{Binney}(2012{\natexlab{b}})}]{Binney2012b}
{Binney} J., 2012{\natexlab{b}}, \mnras, 426, 1328

\bibitem[{{Binney}(2014)}]{Binney2014c}
{Binney} J., 2014, \mnras, 440, 787

\bibitem[{{Binney} \& {Merrifield}(1998)}]{GalacticAstronomy}
{Binney} J., {Merrifield} M., 1998, {Galactic Astronomy}. Princeton University
  Press

\bibitem[{{Binney} et~al.(1997){Binney}, {Gerhard} \& {Spergel}}]{BinneyGS1997}
{Binney} J., {Gerhard} O., {Spergel} D., 1997, \mnras, 288, 365

\bibitem[{{Binney} et~al.(2014{\natexlab{a}})}]{Binney2014b}
{Binney} J. et~al., 2014{\natexlab{a}}, \mnras, 439, 1231

\bibitem[{{Binney} et~al.(2014{\natexlab{b}})}]{Binney2014a}
{Binney} J. et~al., 2014{\natexlab{b}}, \mnras, 437, 351

\bibitem[{{Binney}(2010{\natexlab{b}})}]{BinneyTeneriffe}
{Binney} J.~J., 2010{\natexlab{b}}, in IAC Talks, Astronomy and Astrophysics
  Seminars from the Instituto de Astrof{\'{\i}}sica de Canarias. p. 141

\bibitem[{{Bissantz} \& {Gerhard}(2002)}]{Bissantz2002}
{Bissantz} N., {Gerhard} O., 2002, \mnras, 330, 591

\bibitem[{{Blumenthal} et~al.(1986){Blumenthal}, {Faber}, {Flores} \&
  {Primack}}]{Blumenthal1986}
{Blumenthal} G.~R., {Faber} S.~M., {Flores} R., {Primack} J.~R., 1986, \apj,
  301, 27

\bibitem[{{Bond} et~al.(2010)}]{Bond2010}
{Bond} N.~A. et~al., 2010, \apj, 716, 1

\bibitem[{{Bovy} \& {Rix}(2013)}]{BovyRix2013}
{Bovy} J., {Rix} H.~W., 2013, \apj, 779, 115

\bibitem[{{Bovy} \& {Tremaine}(2012)}]{BovyTremaine2012}
{Bovy} J., {Tremaine} S., 2012, \apj, 756, 89

\bibitem[{{Bovy} et~al.(2012{\natexlab{a}}){Bovy}, {Rix}, {Liu}, {Hogg},
  {Beers} \& {Lee}}]{Bovy2012c}
{Bovy} J., {Rix} H.~W., {Liu} C., {Hogg} D.~W., {Beers} T.~C., {Lee} Y.~S.,
  2012{\natexlab{a}}, \apj, 753, 148

\bibitem[{{Bovy} et~al.(2012{\natexlab{b}})}]{Bovy2012b}
{Bovy} J. et~al., 2012{\natexlab{b}}, \apj, 759, 131

\bibitem[{{Bower} et~al.(2010)}]{Bower2010}
{Bower} R.~G., {Vernon} I., {Goldstein} M., {Benson} A.~J., {Lacey} C.~G.,
  {Baugh} C.~M., {Cole} S., {Frenk} C.~S., 2010, \mnras, 407, 2017

\bibitem[{{Boylan-Kolchin} et~al.(2010){Boylan-Kolchin}, {Springel}, {White} \&
  {Jenkins}}]{BoylanKolchin2010}
{Boylan-Kolchin} M., {Springel} V., {White} S.~D.~M., {Jenkins} A., 2010,
  \mnras, 406, 896

\bibitem[{{Caldwell} \& {Ostriker}(1981)}]{Caldwell1981}
{Caldwell} J.~A.~R., {Ostriker} J.~P., 1981, \apj, 251, 61

\bibitem[{{Das} et~al.(2011){Das}, {Gerhard}, {Mendez}, {Teodorescu} \& {de
  Lorenzi}}]{Das2011}
{Das} P., {Gerhard} O., {Mendez} R.~H., {Teodorescu} A.~M., {de Lorenzi} F.,
  2011, \mnras, 415, 1244

\bibitem[{{Debattista}(2014)}]{Debattista2014}
{Debattista} V.~P., 2014, \mnras, 443, L1

\bibitem[{{Dehnen}(1998)}]{Dehnen1998_UVplane}
{Dehnen} W., 1998, \aj, 115, 2384

\bibitem[{{Dehnen} \& {Binney}(1998)}]{Dehnen1998}
{Dehnen} W., {Binney} J., 1998, \mnras, 294, 429

\bibitem[{{Delfosse} et~al.(2004)}]{Delfosse2004}
{Delfosse} X. et~al., 2004, in R.W. {Hilditch}, H.~{Hensberge}, K.~{Pavlovski},
  eds, Spectroscopically and Spatially Resolving the Components of the Close
  Binary Stars. Astronomical Society of the Pacific Conference Series, Vol.
  318, pp. 166--174

\bibitem[{{Dieterich} et~al.(2012){Dieterich}, {Henry}, {Golimowski}, {Krist}
  \& {Tanner}}]{Dieterich2012}
{Dieterich} S.~B., {Henry} T.~J., {Golimowski} D.~A., {Krist} J.~E., {Tanner}
  A.~M., 2012, \aj, 144, 64

\bibitem[{{Faure} et~al.(2014){Faure}, {Siebert} \& {Famaey}}]{Faure2014}
{Faure} C., {Siebert} A., {Famaey} B., 2014, \mnras, 440, 2564

\bibitem[{{Flynn} et~al.(2006){Flynn}, {Holmberg}, {Portinari}, {Fuchs} \&
  {Jahrei{\ss}}}]{Flynn2006}
{Flynn} C., {Holmberg} J., {Portinari} L., {Fuchs} B., {Jahrei{\ss}} H., 2006,
  \mnras, 372, 1149

\bibitem[{{Garbari} et~al.(2012){Garbari}, {Liu}, {Read} \&
  {Lake}}]{Garbari2012}
{Garbari} S., {Liu} C., {Read} J.~I., {Lake} G., 2012, \mnras, 425, 1445

\bibitem[{{Gillessen} et~al.(2009)}]{Gillessen2009b}
{Gillessen} S., {Eisenhauer} F., {Fritz} T.~K., {Bartko} H., {Dodds-Eden} K.,
  {Pfuhl} O., {Ott} T., {Genzel} R., 2009, \apjl, 707, L114

\bibitem[{{Gilmore} \& {Reid}(1983)}]{Gilmore1983}
{Gilmore} G., {Reid} N., 1983, \mnras, 202, 1025

\bibitem[{{Gnedin} et~al.(2004){Gnedin}, {Kravtsov}, {Klypin} \&
  {Nagai}}]{Gnedin2004}
{Gnedin} O.~Y., {Kravtsov} A.~V., {Klypin} A.~A., {Nagai} D., 2004, \apj, 616,
  16

\bibitem[{{Gnedin} et~al.(2010){Gnedin}, {Brown}, {Geller} \&
  {Kenyon}}]{Gnedin2010}
{Gnedin} O.~Y., {Brown} W.~R., {Geller} M.~J., {Kenyon} S.~J., 2010, \apjl,
  720, L108

\bibitem[{{Guo} et~al.(2010){Guo}, {White}, {Li} \& {Boylan-Kolchin}}]{Guo2010}
{Guo} Q., {White} S., {Li} C., {Boylan-Kolchin} M., 2010, \mnras, 404, 1111

\bibitem[{{Hinshaw} et~al.(2013)}]{WMAP-9}
{Hinshaw} G. et~al., 2013, \apjs, 208, 19

\bibitem[{{Holmberg} \& {Flynn}(2004)}]{Holmberg2004}
{Holmberg} J., {Flynn} C., 2004, \mnras, 352, 440

\bibitem[{{Holmberg} et~al.(2007){Holmberg}, {Nordstr{\"o}m} \&
  {Andersen}}]{Holmberg2007}
{Holmberg} J., {Nordstr{\"o}m} B., {Andersen} J., 2007, \aap, 475, 519

\bibitem[{{Iocco} et~al.(2011){Iocco}, {Pato}, {Bertone} \&
  {Jetzer}}]{Iocco2011}
{Iocco} F., {Pato} M., {Bertone} G., {Jetzer} P., 2011, \jcap, 11, 029

\bibitem[{{Juri{\'c}} et~al.(2008)}]{Juric2008}
{Juri{\'c}} M. et~al., 2008, \apj, 673, 864

\bibitem[{{Kahn} \& {Woltjer}(1959)}]{Kahn1959}
{Kahn} F.~D., {Woltjer} L., 1959, \apj, 130, 705

\bibitem[{{Kent} et~al.(1991){Kent}, {Dame} \& {Fazio}}]{Kent1991}
{Kent} S.~M., {Dame} T.~M., {Fazio} G., 1991, \apj, 378, 131

\bibitem[{{Klypin} et~al.(1999){Klypin}, {Kravtsov}, {Valenzuela} \&
  {Prada}}]{Klypin1999}
{Klypin} A., {Kravtsov} A.~V., {Valenzuela} O., {Prada} F., 1999, \apj, 522, 82

\bibitem[{{Kneib} \& {Natarajan}(2011)}]{Kneib2011}
{Kneib} J.~P., {Natarajan} P., 2011, \aapr, 19, 47

\bibitem[{{Korchagin} et~al.(2003){Korchagin}, {Girard}, {Borkova}, {Dinescu}
  \& {van Altena}}]{Korchagin2003}
{Korchagin} V.~I., {Girard} T.~M., {Borkova} T.~V., {Dinescu} D.~I., {van
  Altena} W.~F., 2003, \aj, 126, 2896

\bibitem[{{Kordopatis} et~al.(2013)}]{RAVE_DR4}
{Kordopatis} G. et~al., 2013, \aj, 146, 134

\bibitem[{{Kuijken} \& {Gilmore}(1991)}]{Kuijken1991}
{Kuijken} K., {Gilmore} G., 1991, \apjl, 367, L9

\bibitem[{{Lee} et~al.(2011)}]{Lee2011b}
{Lee} Y.~S. et~al., 2011, \apj, 738, 187

\bibitem[{{Li} \& {White}(2008)}]{Li2008}
{Li} Y.~S., {White} S.~D.~M., 2008, \mnras, 384, 1459

\bibitem[{{Macci{\`o}} et~al.(2008){Macci{\`o}}, {Dutton} \& {van den
  Bosch}}]{Maccio2008}
{Macci{\`o}} A.~V., {Dutton} A.~A., {van den Bosch} F.~C., 2008, \mnras, 391,
  1940

\bibitem[{{Malhotra}(1995)}]{Malhotra1995}
{Malhotra} S., 1995, \apj, 448, 138

\bibitem[{{Marinacci} et~al.(2010){Marinacci}, {Binney}, {Fraternali},
  {Nipoti}, {Ciotti} \& {Londrillo}}]{Marinacci2010}
{Marinacci} F., {Binney} J., {Fraternali} F., {Nipoti} C., {Ciotti} L.,
  {Londrillo} P., 2010, \mnras, 404, 1464

\bibitem[{{Martinsson} et~al.(2013){Martinsson}, {Verheijen}, {Westfall},
  {Bershady}, {Andersen} \& {Swaters}}]{Martinsson2013}
{Martinsson} T.~P.~K., {Verheijen} M.~A.~W., {Westfall} K.~B., {Bershady}
  M.~A., {Andersen} D.~R., {Swaters} R.~A., 2013, \aap, 557, A131

\bibitem[{{McMillan}(2011)}]{McMillan2011}
{McMillan} P.~J., 2011, \mnras, 414, 2446

\bibitem[{{McMillan} \& {Binney}(2010)}]{McMillan2010}
{McMillan} P.~J., {Binney} J.~J., 2010, \mnras, 402, 934

\bibitem[{{McMillan} \& {Binney}(2013)}]{McMillan2013}
{McMillan} P.~J., {Binney} J.~J., 2013, \mnras, 433, 1411

\bibitem[{{Mirabolfathi}(2013)}]{Mirabolfathi}
{Mirabolfathi} N., 2013, ArXiv e-prints, 1308.0044

\bibitem[{{Moni Bidin} et~al.(2012{\natexlab{a}}){Moni Bidin}, {Carraro} \&
  {M{\'e}ndez}}]{MoniBidin2012a}
{Moni Bidin} C., {Carraro} G., {M{\'e}ndez} R.~A., 2012{\natexlab{a}}, \apj,
  747, 101

\bibitem[{{Moni Bidin} et~al.(2012{\natexlab{b}}){Moni Bidin}, {Carraro},
  {M{\'e}ndez} \& {Smith}}]{MoniBidin2012b}
{Moni Bidin} C., {Carraro} G., {M{\'e}ndez} R.~A., {Smith} R.,
  2012{\natexlab{b}}, \apj, 751, 30

\bibitem[{{Navarro} \& {Steinmetz}(2000)}]{Navarro2000a}
{Navarro} J.~F., {Steinmetz} M., 2000, \apj, 528, 607

\bibitem[{{Navarro} et~al.(1996){Navarro}, {Frenk} \& {White}}]{NFW1996}
{Navarro} J.~F., {Frenk} C.~S., {White} S.~D.~M., 1996, \apj, 462, 563

\bibitem[{{Nesti} \& {Salucci}(2013)}]{Nesti2013}
{Nesti} F., {Salucci} P., 2013, \jcap, 7, 016

\bibitem[{{Nordstr{\"o}m} et~al.(2004)}]{GCS2004}
{Nordstr{\"o}m} B. et~al., 2004, \aap, 418, 989

\bibitem[{{Piffl} et~al.(2014)}]{Piffl2014}
{Piffl} T. et~al., 2014, \aap, 562, A91

\bibitem[{{Planck Collaboration} et~al.(2013)}]{Planck2013-XVI}
{Planck Collaboration} et~al., 2013, ArXiv e-prints, 1303.5076

\bibitem[{Press et~al.(2007)Press, Teukolsky, Vetterling \&
  Flannery}]{NumericalRecipes}
Press W.~H., Teukolsky S.~A., Vetterling W.~T., Flannery B.~P., 2007, Numerical
  Recipes 3rd Edition: The Art of Scientific Computing. Cambridge University
  Press, New York, NY, USA, 3rd ed.

\bibitem[{{Read}(2014)}]{Read2014}
{Read} J.~I., 2014, Journal of Physics G Nuclear Physics, 41, 063101

\bibitem[{{Reid} \& {Brunthaler}(2004)}]{Reid2004}
{Reid} M.~J., {Brunthaler} A., 2004, \apj, 616, 872

\bibitem[{{Reid} et~al.(2014)}]{Reid2014}
{Reid} M.~J. et~al., 2014, \apj, 783, 130

\bibitem[{{Sackett}(1997)}]{Sackett1997}
{Sackett} P.~D., 1997, \apj, 483, 103

\bibitem[{{Salucci} et~al.(2010){Salucci}, {Nesti}, {Gentile} \& {Frigerio
  Martins}}]{Salucci2010}
{Salucci} P., {Nesti} F., {Gentile} G., {Frigerio Martins} C., 2010, \aap, 523,
  A83

\bibitem[{{Sanders}(2012)}]{Sanders2012}
{Sanders} J.~L., 2012, \mnras, 425, 2228

\bibitem[{{Sch{\"o}nrich}(2012)}]{Schoenrich2012}
{Sch{\"o}nrich} R., 2012, \mnras, 427, 274

\bibitem[{{Sch{\"o}nrich} \& {Bergemann}(2014)}]{Schoenrich2014b}
{Sch{\"o}nrich} R., {Bergemann} M., 2014, \mnras, 443, 698

\bibitem[{{Sch{\"o}nrich} et~al.(2010){Sch{\"o}nrich}, {Binney} \&
  {Dehnen}}]{Schoenrich2010}
{Sch{\"o}nrich} R., {Binney} J., {Dehnen} W., 2010, \mnras, 403, 1829

\bibitem[{{Siebert} et~al.(2012)}]{Siebert2012}
{Siebert} A. et~al., 2012, \mnras, 425, 2335

\bibitem[{{Springel} \& {White}(1999)}]{SW1999}
{Springel} V., {White} S.~D.~M., 1999, \mnras, 307, 162

\bibitem[{{Steinmetz} et~al.(2006)}]{RAVE_DR1}
{Steinmetz} M. et~al., 2006, \aj, 132, 1645

\bibitem[{{van Albada} et~al.(1985){van Albada}, {Bahcall}, {Begeman} \&
  {Sancisi}}]{vanAlbada1985}
{van Albada} T.~S., {Bahcall} J.~N., {Begeman} K., {Sancisi} R., 1985, \apj,
  295, 305

\bibitem[{{van der Kruit} \& {Searle}(1981)}]{vanderKruit1981}
{van der Kruit} P.~C., {Searle} L., 1981, \aap, 95, 105

\bibitem[{{Velander} et~al.(2014)}]{Velander2014}
{Velander} M. et~al., 2014, \mnras, 437, 2111

\bibitem[{{Wilkinson} \& {Evans}(1999)}]{Wilkinson1999}
{Wilkinson} M.~I., {Evans} N.~W., 1999, \mnras, 310, 645

\bibitem[{{Williams} et~al.(2013)}]{Williams2013}
{Williams} M.~E.~K. et~al., 2013, \mnras

\bibitem[{{Yanny} et~al.(2009)}]{SEGUEpaper}
{Yanny} B. et~al., 2009, \aj, 137, 4377

\bibitem[{{York} et~al.(2000)}]{York2000}
{York} D.~G. et~al., 2000, \aj, 120, 1579

\bibitem[{{Zacharias} et~al.(2013){Zacharias}, {Finch}, {Girard}, {Henden},
  {Bartlett}, {Monet} \& {Zacharias}}]{UCAC4_paper}
{Zacharias} N., {Finch} C.~T., {Girard} T.~M., {Henden} A., {Bartlett} J.~L.,
  {Monet} D.~G., {Zacharias} M.~I., 2013, \aj, 145, 44

\bibitem[{{Zhang} et~al.(2013){Zhang}, {Rix}, {van de Ven}, {Bovy}, {Liu} \&
  {Zhao}}]{Zhang2013}
{Zhang} L., {Rix} H.~W., {van de Ven} G., {Bovy} J., {Liu} C., {Zhao} G., 2013,
  \apj, 772, 108

\end{thebibliography}
\label{lastpage}
\end{document}